\documentclass[acmtog]{acmart}
\acmSubmissionID{257}

\usepackage{amsmath, amsthm, amsfonts}
\usepackage{graphicx}
\usepackage{multirow}
\usepackage{color}
\usepackage{enumitem}

\definecolor{green}{rgb}{0, 0.5, 0}
\definecolor{orange}{rgb}{0.6, 0.3, 0.1}
\definecolor{red}{rgb}{1.0, 0.0, 0.0}
\definecolor{teal}{rgb}{0.0, 0.4, 0.4}
\definecolor{purple}{rgb}{0.65,0,0.65}
\definecolor{saffron}{rgb}{0.8,0.55,0.1}
\definecolor{turquoise}{rgb}{0.0,0.5,0.5}
\definecolor{brown}{rgb}{0.5, 0.16, 0.16}
\definecolor{brickred}{rgb}{.6, .2 .1}
\definecolor{coral}{rgb}{1,0.45,0.33}
\definecolor{newcolor}{rgb}{.8,.349,.1}

\newcommand{\jw}[1]{{\color{black}#1}}

\newcommand{\wdx}[1]{{\color{black}#1}}

\newcommand{\hq}[1]{{\color{black}#1}}

\newcommand{\eg}{{\textit{e.g., }}}
\newcommand{\ie}{{\textit{i.e., }}}

\newcommand{\ignore}[1]{}
\usepackage{subfigure}

\usepackage{wrapfig}
\usepackage{numprint}
\npthousandsep{,}

\begin{document}
	
\title{TwinTex: Geometry-aware Texture Generation for Abstracted 3D Architectural Models}	

\author{Weidan Xiong}
\email{xiongweidan@gmail.com}
\affiliation{%
	\institution{Shenzhen University}
	\country{China}	
}

\author{Hongqian Zhang}
\affiliation{%
	\institution{Shenzhen University}
	\country{China}	
}

\author{Botao Peng}
\affiliation{%
	\institution{Shenzhen University}
	\country{China}	
}

\author{Ziyu Hu}
\affiliation{%
	\institution{Guangdong Artificial Intelligence and Digital Economy Laboratory (SZ), Shenzhen University}
	\country{China}	
}

\author{Yongli Wu}
\affiliation{%
	\institution{Guangdong Artificial Intelligence and Digital Economy Laboratory (SZ), Shenzhen University}
	\country{China}	
}

\author{Jianwei Guo}
\affiliation{%
	\institution{MAIS, Institute of Automation, Chinese Academy of Sciences}
	\country{China}	
}

\author{Hui Huang}
\email{hhzhiyan@gmail.com}
\authornote{Corresponding author: Hui Huang (hhzhiyan@gmail.com)}
\affiliation{
	\institution{Shenzhen University}
	\department{College of Computer Science \& Software Engineering}
	\country{China}
}

\renewcommand\shortauthors{W. Xiong, H. Zhang, B. Peng, Z. Hu, Y. Wu, J. Guo, and H. Huang}

\begin{abstract}
Coarse architectural models are often generated at scales ranging from individual buildings to scenes for downstream applications such as Digital Twin City, Metaverse, LODs, etc. Such piece-wise planar models can be abstracted as twins from 3D dense reconstructions. However, these models typically lack realistic texture relative to the real building or scene, making them unsuitable for vivid display or direct reference. In this paper, we present \emph{TwinTex}, the first automatic texture mapping framework to generate a photo-realistic texture for a piece-wise planar proxy.
Our method addresses most challenges occurring in such twin texture generation. Specifically, for each primitive plane, we first select a small set of photos with greedy heuristics considering photometric quality, perspective quality and facade texture completeness. Then, different levels of line features (LoLs) are extracted from the set of selected photos to generate guidance for later steps. With LoLs, we employ optimization algorithms to align texture with geometry from local to global.
Finally, we fine-tune a diffusion model with a multi-mask initialization component and a new dataset to inpaint the missing region. Experimental results on many buildings, indoor scenes and man-made objects of varying complexity demonstrate the generalization ability of our algorithm.
Our approach surpasses state-of-the-art texture mapping methods in terms of high-fidelity quality and reaches a human-expert production level with much less effort.
\end{abstract}


\begin{teaserfigure}
  \centering
  \includegraphics[width=\linewidth]{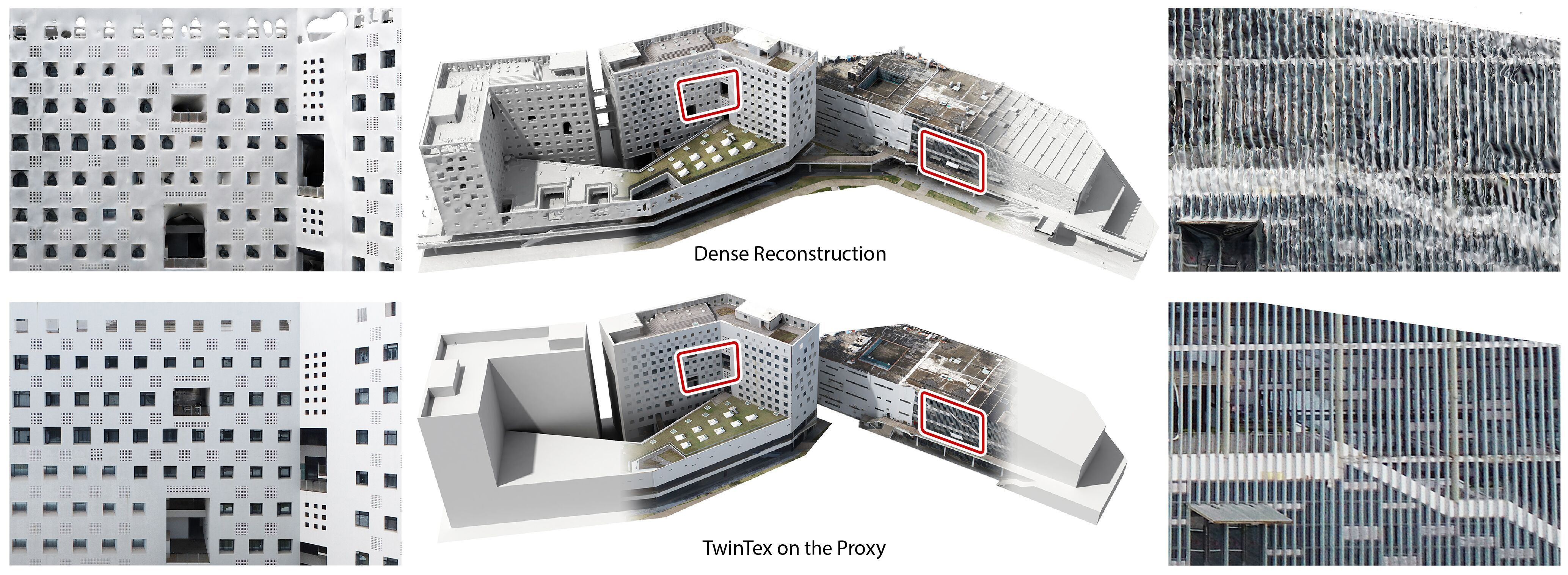}
  \caption{We introduce \textbf{TwinTex}, a fully automatic method to produce a detailed texture map with high photometric and perceptual quality for a piece-wise planar 3D proxy mode. Such a proxy (\numprint{1489} vertices, 739 faces, shown in the bottom-middle) is an abstracted version of a 3D dense model (\numprint{656571} vertices, \numprint{1313550} faces, shown at the top-middle), which is reconstructed from multi-view aerial images captured by a drone.
  }
  \label{fig:teaser}
\end{teaserfigure}

\setcopyright{acmlicensed}
\acmJournal{TOG}
\acmYear{2023} \acmVolume{42} \acmNumber{6} \acmArticle{} \acmMonth{12} \acmPrice{15.00}\acmDOI{10.1145/3618328}

\maketitle

\section{Introduction}
\label{sec:intro}

\begin{figure}
	\centering
	\includegraphics[width=\linewidth]{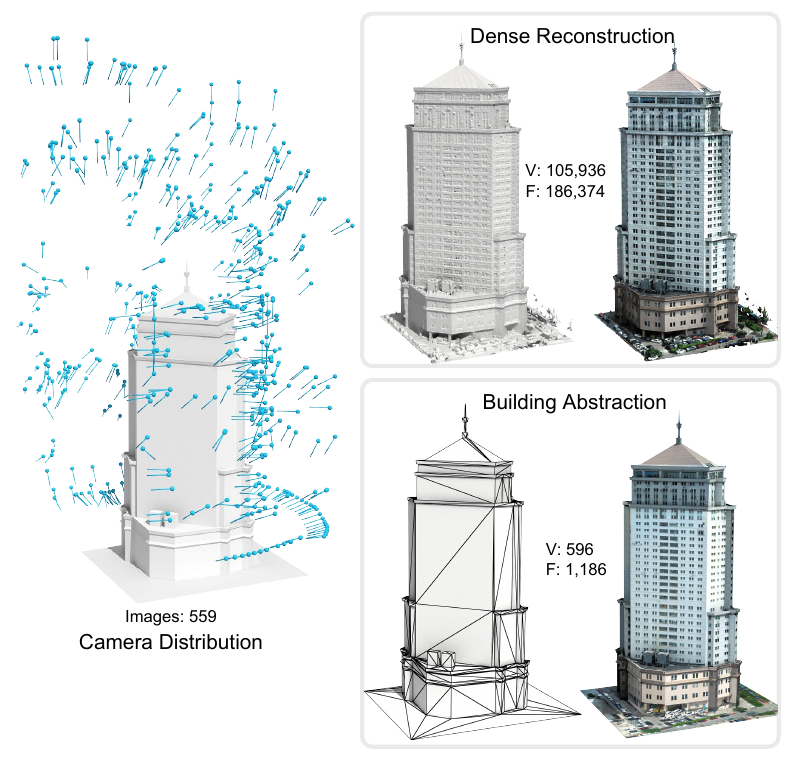}
	\caption{Left: Aerial distribution of the drone that collected the input photos (559 images). Top right: 3D dense reconstruction of a Headquarter. Bottom right: An abstracted twin model with a much smoother appearance and smaller storage needs.}
	\label{fig:abstraction}
\end{figure}

With the advances in active sensors (\eg LiDAR), fully automated \emph{Multiple View Stereo} (MVS) and aerial-based photogrammetry sensing technology, many mature algorithms have been developed to generate complete and high-fidelity 3D dense reconstruction~\cite{musialski2013survey,Smith2018Aerial,DroneScan20} with textures~\cite{waechter2014let}. Architectural models with simplified topology and twin shapes of such dense models can be generated from procedural modeling, approximated from dense reconstruction, or created manually by artists and architects. These abstracted twin models are tractable in plenty of downstream applications due to their sharper features, much lower needs on storage, \ignore{more intuitive in editing, }and more suitable for transmission and real-time rendering, as shown in Fig.~\ref{fig:abstraction}.

\begin{figure}
	\centering
	\includegraphics[width=\linewidth]{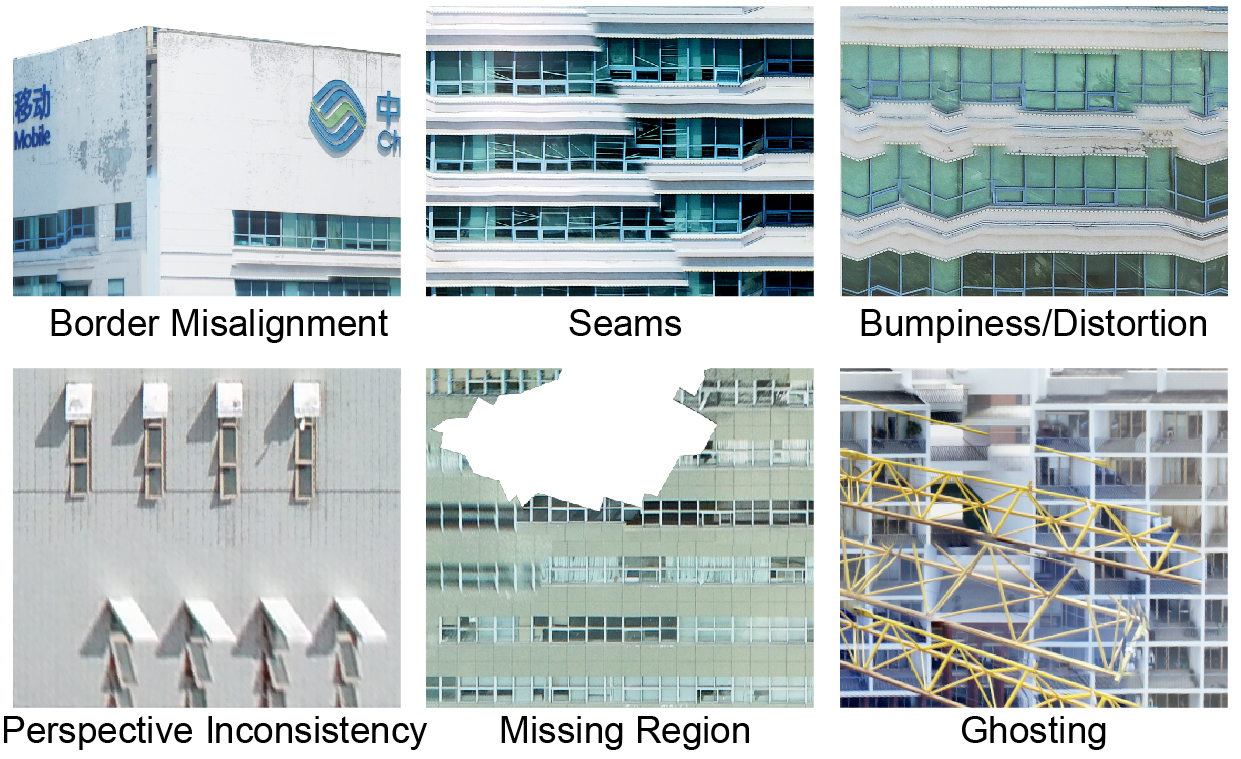}
	\caption{Illustration of several typical issues occurred in texturing architectural proxies.}
	\label{fig:texturing-issues}
\end{figure}

However, the texture information contained in the original dense reconstruction is often not maintained after abstraction. Thus, with all the calibrated RGB images, how to texture the twin proxy for the level of realism then becomes a challenging problem.
An intuitive way is to employ image-based texture generation methods ~\cite{lempitsky2007seamless,waechter2014let,zhou2014color}. However, these approaches would produce distorted and misaligned artifacts as illustrated in Fig.~\ref{fig:texturing-issues}.
Although patch-based optimization method~\cite{bi2017patch} can reduce such artifacts to some extent by synthesizing photometrically-consistent aligned images to correct misalignments, it would possibly deliver blurring textures or missing regions. These issues are mainly caused by: \romannumeral1) the inaccurate camera parameters, \romannumeral2) the significant geometric differences between the twin mesh and ground-truth models brought by the abstraction process, and \romannumeral3) the large difference among viewing angles. The combination of these cases makes our problem even more challenging and complicated.
The urban proxy models often exhibit large-scale planar structures. In practice, the modeling artists often manually create high-quality textures plane-by-plane.
This is an extremely tedious, time-consuming, and labor-intensive process. It took quite a few days for a professional modeler to texture an abstracted model such as the Polytech shown in Fig.~\ref{fig:teaser} with real photos. Texturing a model with a higher level of detail would take much longer.

In this paper, we propose a plane-based methodology and fully automated algorithms to generate high-quality texture maps for piece-wise planar architectural models given a large set of unordered RGB photos. Our method addresses most challenges occurring in such twin texture generation: the large number and high resolution of input images, their drastically varying properties such as photometric quality, perspective quality (large variation of viewing angles), heavy occludes (\eg trees and buildings), and texture-geometry misalignment (introduced by incorrect camera parameters and model abstraction process).
Given a coarse model, our algorithm decomposes it into a set of planar polygonal shapes, for each of which we aim at generating a high-quality and complete texture map. We first present a plane-oriented view selection approach to select views that best cover each planar shape with high photometric and perspective consistency. Based on the line segments extracted from each view, a line-guided texture stitching is introduced to create a single texture map to preserve geometric features. Finally, we perform texture optimization to improve the illumination consistency and fill the missing regions in the texture map.
Experiments are performed to show the effectiveness and robustness of our method.
In summary, our work makes the following contributions:
\begin{itemize}
\item A fully automatic \textbf{TwinTex}, which produces high-quality and geometry-aware texture for a simplified proxy model.

\item A view selection algorithm that can select a small number of candidate photos for each primitive plane considering texture completeness, perspective and photometric quality.

\item A plane-based texture mapping methodology via geometry-aware image stitching guided by LoLs.

\item An architectural dataset including various scenes and building components. We utilize this dataset and fine-tune a diffusion probabilistic model with a novel multi-mask initialization component to complete the texture maps.

\end{itemize}

\section{Related Work}\label{sec:rw}

\subsection{Image-based Texture Mapping}

Texture mapping is an important approach for authoring photo-realistic 3D objects without increasing their geometric complexity~\cite{yuksel2019rethinking}.
View-dependent texture mapping works~\cite{debevec1996modeling,debevec1998efficient} composite multiple views of a scene by assuming the existence of a global mesh and storing all appearance variation in textures.
Although they also use a view selection procedure to project a single image onto the model and then merge several image projections, they only determine for each polygon in the model in which photos it is visible.
Instead, we present a novel view selection algorithm that considers texture completeness, perspective quality, and photometric quality.

To generate a single consistent texture map, early works on image-based texture mapping usually select multiple images for each face and blend them into textures by using different weighting strategies~\cite{bernardini2001high, callieri2008masked}. However, such methods could generate blurry and ghosting results due to camera calibration errors and inaccurate geometry.
To reduce those artifacts caused by multi-image blending, later approaches choose to select only one view per face, then conduct seam optimization to create a texture patch and to avoid visible seams between adjacent patches~\cite{lempitsky2007seamless,wang2018seamless,gal2010seamless, waechter2014let,fu2018texture,fu2020joint}. They typically solve a discrete labeling problem by using different data terms and smoothness terms to construct the Markov random field to select an optimal  texture image for each face.
Whereas, these approaches still suffer from misaligned seams in challenging cases.
Another category of warping-based methods jointly rectifies the camera poses and geometric errors. Zhou and Koltun~\shortcite{zhou2014color} use local image warping to optimize camera poses in tandem with non-rigid correction functions for all images. This approach is then extended by~\cite{bi2017patch} to propose a patch-based optimization that handles larger geometry inaccuracies by correctly aligning the input images. This method estimates the bidirectional similarity of different images, which suffers from high computational costs.

Although high-quality texture maps can be generated from the above methods, the resultant quality of texture mapping still greatly depends on the quality of reconstruction and accuracy of camera calibration. None of them can be naively applied to generate the realistic texture of a large-scale structured building or scene. Firstly, the abstracted proxy model has significant structural differences from the original dense mesh. Moreover, constrained by the complicated environment, the collected aerial data often has highly uneven camera distribution, large camera calibration errors and large variations in viewing angles. Current methods still cannot handle these problems well, resulting in obvious artifacts. Instead, we choose to generate texture in a plane-based manner that is more intuitive and human alike.

\begin{figure*}[tp]
	\centering
	\includegraphics[width=\linewidth]{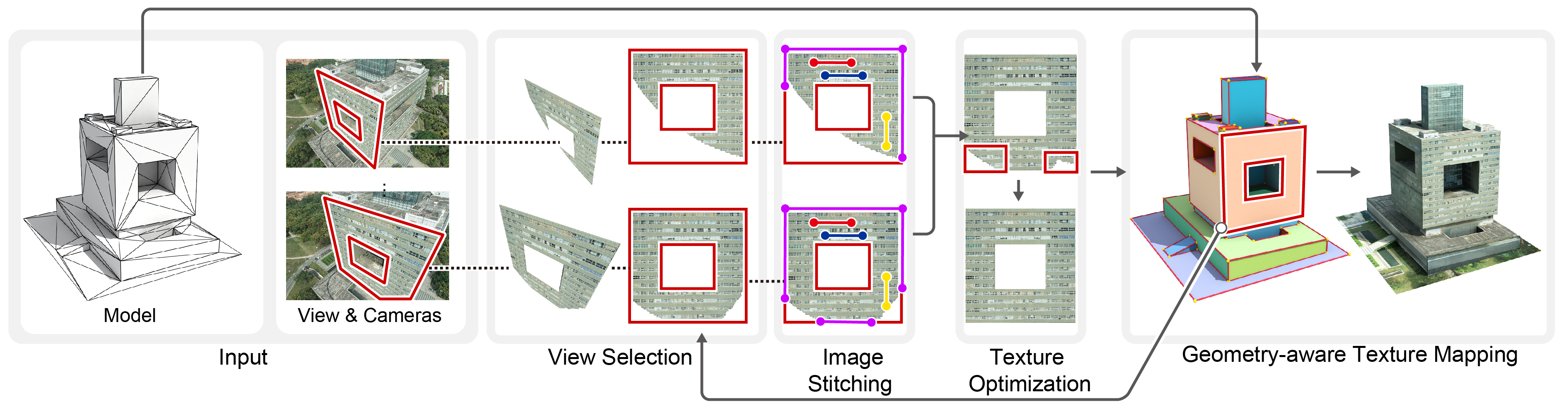}
	\caption{Overview of our algorithm on the Hitech building. Taking an abstracted model and the UAV images as input, our framework detects the proxy polygons in the mesh and generates one stitched texture map for each proxy polygon. Firstly, a small subset of views is selected from all the visible views for each proxy polygon. Then, all the selected views are stitched together to form an enlarged texture map. Note that important geometric primitives such as line segments contained in the original images are well preserved after the image stitching. Finally, the texture map is optimized by performing global consistency across all the visible regions and inpainting the incomplete region of the underlying planar shape.}
	\label{fig:pipeline}
\end{figure*}

\subsection{Structure-aware Scene Reconstruction and Texturing}

Creating structured models from noisy raw data has become an ever-increasing demand in urban digitization. \jw{\citet{ceylan2012factored,ceylan2014coupled} propose image-based 3D urban building reconstruction frameworks by exploiting the presence of linear and symmetric features (\eg lines, repeated elements) in facade for image registration and 3D reconstruction.
To generate simplified meshes,} a popular way is to first reconstruct a 3D dense mesh and then simplify it using geometry simplification~\cite{garland1997surface} or piece-wise planar structure approximation methods~\cite{cohen2004variational, verdie2015lod, salinas2015structure}. Another way is to directly detect a set of planar primitives from the point clouds, then explore the arrangements of primitives to provide a compact polygonal mesh~\cite{Monszpart2015rapter, Nan2017polyfit, fang2020connect, bauchet2020kinetic, bouzas2020structure, Pan2022Efficient, Guo2022}. In practice, such proxy models are usually generated manually or by procedural modeling~\cite{sinha2008interactive} with arbitrary topology (\eg manifolds or non-manifolds).

However, texturing such abstracted building models has drawn less attention. Several methods take into account the generation of texture maps while conducting a lightweight geometric reconstruction~\cite{sinha2008interactive, garcia2013automatic, huang20173dlite, maier2017intrinsic3d, wang2018plane}. For example, \citet{garcia2013automatic} computes a per-building volumetric proxy, then uses a surface graph-cut method to stitch aerial images and yield a visually coherent texture map. However, their reconstruction results are 2.5D which can only restore textures under limited perspectives, losing important structures.
Recently, deep learning approaches based on image-image translation networks have been proposed to synthesize texture details for coarse meshes of urban areas~\cite{Kelly2018FrankenGAN, georgiou2021projective}. Although data-driven methods are able to generate various styles of textures, they require extensive training sets, and the synthesized textures lack realism comparing to the original real scenes.
The technique of NeRF~\cite{mueller2022instant} shows great ability to generate high-quality 3D shapes and textures~\cite{metzer2022latent,baatz2022nerf}. However, remember that we aim at very high-resolution input and output images which require extremely large memory, employing a small set of input images is not enough for NeRF-based methods to generate satisfying results, which makes them less applicable in our scenario.

\subsection{Image Stitching and Completion}

Image stitching is to combine multiple images with overlapping sections to produce a single panoramic or high-resolution image~\cite{szeliski2007image, Mehta2011}.
Early methods estimate an optimal global transformation for each input image~\cite{brown2007automatic}. However, they only work well for scenes near planar or with slight view disparity, and usually generate ghosting artifacts or projective distortion for general scenes.
For more accurate stitching, many approaches adopt spatially-varying warps to process different regions of an image, including smoothly varying affine transformation~\cite{lin2011smoothly}, dual or quasi homographies~\cite{gao2011constructing, li2017quasi}, shape-preserving half-projective warps~\cite{chang2014shape, lin2015adaptive}, and smoothly varying homography field~\cite{zaragoza2013projective}, etc.
\jw{Aiming at building facades, \citet{musialski2010interactive} present a system for generating approximately orthographic facade textures from a set of perspective photographs. They perform multi-view image stitching over planar proxies by using a gradient-domain stitching to avoid visible seams.}

However, they are usually time-consuming and unsuitable for scenes with large parallax and view disparity. Several methods are then proposed to obtain better alignment with less distortion, especially for large parallax~\cite{zhang2016multi, lee2020warping}, but linear structures in the image are not well maintained. After that, instead of just using point feature matching, line features are introduced to preserve both linear structures~\cite{li2015dual, xiang2018image, liao2019single, jia2021leveraging}.
All these methods apply a uniform grid to warp the images, which are not flexible enough to align the regions with dense geometric features. Unlike them, we propose to use an adaptive grid to locally warp the images and better align the large-scale line features.

Image completion, also known as image inpainting, is the process of restoring missing or damaged areas in an image. Extensive research has been conducted on image completion, and existing methods can be classified into three types: 1) diffusion-based methods, 2) patch-based methods and 3) deep learning methods based on generative models. A detailed review is out of the scope of this paper, and we refer the reader to recent works~\cite{dhariwal2021diffusion, lugmayr2022repaint} and surveys~\cite{guillemot2014image, rojas2020a, elharrouss2020image, jam2021a}.

\section{Problem Statement and Overview}
\label{sec:overview}

The objective of our approach is to automatically generate realistic and geometry-aware texture maps for a piece-wise planar model, \ie urban building or scene.
Our algorithm takes a set of calibrated camera views $\{ \mathcal{I} \}$ and an abstracted model $\mathcal{M}$ as input, and outputs high-quality texture maps. 
The proxy model $\mathcal{M}$ is a polygonal mesh consisting of a set of planar polygonal shapes $\mathcal{P}_i$, each of which is referred to as a \emph{proxy polygon}.

To produce a texture map for each proxy polygon $\mathcal{P}_i$, the first step should be to select one or multiple images as candidate maps for $\mathcal{P}_i$. However, unlike previous texturing problems dealing with dense models (small triangles), it is usual that none of the given images can cover such large proxy polygons in abstracted versions. Hence, we employ the latter scheme and select multiple images for each polygon shape.
However, there are still several challenges that exist in our scenario: i) Photometric issue. There are seams, illumination resolution inconsistency between adjacent patches. For large scale scenes, the frequent ghosting effect is mainly introduced by the dynamic instances, such as cars and tower crane. ii) Perspective issue. Even with photometric consistent patches, it would still be apparent that the stitched texture comes from several images with largely varied viewing angles. iii) Image-to-image and image-to-geometry misalignment. This is introduced by both the inaccuracy of camera parameters and the simplification process. iv) Facade incompleteness. In the absence of semantic guidance, how to fill a largely empty region with geometrically and semantically consistent content remains a problem.

\begin{figure}
	\centering
	\includegraphics[width=.75\linewidth]{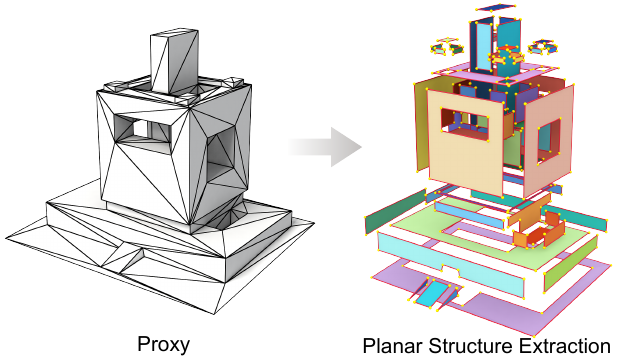}
	\caption{\wdx{The input mesh $\mathcal{M}$ is first segmented into a set of planar regions represented with proxy polygons $\{\mathcal{P}_i\}$. The shape of each proxy polygon is then extracted as proxy boundary $\mathcal{B}_i$ (visualized with red border lines).}}
	\label{fig:boundary_extraction}
\end{figure}

To address the above problems, we propose a plane-based and geometry-aware texture mapping methodology, as shown in Fig.~\ref{fig:pipeline}.
First, we extract high-quality views and visible regions for each proxy polygon $\mathcal{P}_i$. An effective greedy heuristic method is proposed to select a small set of views $\{\mathcal{I}\}_i$ that best cover each $\mathcal{P}_i$ with high photometric and perspective consistency (Sec.~\ref{sec:view-selection}).
Next, in Sec.~\ref{sec:LOLs}, the rich line features contained in the selected images $\{\mathcal{I}\}_i$ are detected and organized into Level of Lines (LoLs).
After that, all the selected images are sequentially warped to their relative proxy boundary based on LoLs and stitched into a single texture map. Sec.~\ref{sec:texture-stitching} provides details.
Then, each texture map is optimized by a series of operators to reveal illumination consistency. Finally, we provide an architectural image dataset and train a multi-mask Diffusion Model to inpaint the missing regions in texture maps, see Sec.~\ref{sec:texture-optimization}.
Our algorithm has been implemented as a customized plug-in of a widely used graphics engine called Houdini\footnote{https://www.sidefx.com/}.

\section{Plane-oriented View Selection}
\label{sec:view-selection}

\paragraph{Proxy polygon extraction.}
The input proxy $\mathcal{M=(E, F)}$ can be created manually, generated procedurally, or abstracted from a dense reconstruction. 
$\mathcal{M}$ is a polygonal mesh where each basic element of $\mathcal{E}$ is an edge and $\mathcal{F}$ is either a polygonal or triangular facet, as shown in Fig.~\ref{fig:boundary_extraction}. We first segment $\mathcal{M}$ into a set of planar polygonal shapes, each of which is represented as a \emph{proxy polygon} $\mathcal{P}_i$. We input $\mathcal{P}_i$ by utilizing a region-growing algorithm based on face normal. We then detect the corresponding proxy boundary $\mathcal{B}_i$ by choosing and sorting the boundary edges of $\mathcal{P}_i$. Fig.~\ref{fig:boundary_extraction} shows the result of this step on the scene of Hitech.

\textbf{View selection.}
Given a large set of input calibrated photos, the goal of this step is to select a small set of camera views ${ \{ \mathcal{I} \} }_{i}$ as candidate feature maps to best cover each proxy polygon $\mathcal{P}_i$.
We first apply frustum-culling and visibility detection to filter out the views and pixels which are invisible, far away from $\mathcal{P}_i$, or have extremely inclined viewing angles. The filtered images are then projected onto $\mathcal{P}_i$, and the projected regions $\{ \mathcal{R} \}$ covering the proxy polygon are extracted as candidate images, as shown in Fig.~\ref{fig:pipeline}.

\begin{figure}[!t]
	\centering
	\includegraphics[width=\linewidth]{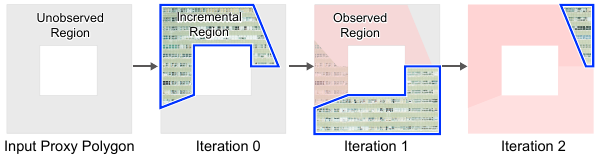}
	\caption{The view selection process of a proxy polygon $\mathcal{P}_i$. The red polygons denote the observed regions, while the grey ones denote the unobserved regions. The polygon with blue outline denotes the incremental region ${A_i^j}$ of selected image $\mathcal{R}_{j}$ in the current iteration. }
	\label{fig:selection}
\end{figure}

Next, we propose an iterative view selection algorithm to select a smaller set of candidate views from the projected images covering $\mathcal{P}_i$ (see Fig.~\ref{fig:selection}).
Starting from an empty set $\{\mathcal{I} \}_{i}^{0}=\emptyset$, we select the best projected image ${\mathcal{R}_j}$ with the highest quality (denoted by $Q_j$) in each iteration $k$ to create $\{ \mathcal{I} \}_{i}^{k} = \{ \mathcal{I} \}_{i}^{k-1} \cup R_j$.
This process converges when there is no projected image left to examine, or the $\{ \mathcal{I} \}_{i}^{k}$ can fully cover $\mathcal{P}_i$, i.e., the area of unobserved regions in $\mathcal{P}_i$ is smaller than a given threshold $\tau$.
We measure the quality $Q_j$ of each ${\mathcal{R}_j}$ by its incremental coverage ratio of the proxy polygon, its photometric and perspective quality and consistency with $\{\mathcal{I} \}_{i}^{k-1}$.
The score $Q_j$ is calculated as the weighted sum of two terms:
\begin{equation}
	Q_j = Q_{photo} + {\lambda}_p Q_{persp} ,
\end{equation}
where $Q_{photo}$ and $Q_{persp}$ represent the photometric and perspective quality, respectively. $\lambda_p$ is the weight of $Q_{persp}$.

\paragraph{Photometric quality.} This term favors $\mathcal{R}_{j}$ with large incremental projection region, high resolution, and high photometric similarity to all the projected images and previously selected images. In detail, the coverage ratio of $\mathcal{R}_{j}$ against proxy polygon $\mathcal{P}_i$ at iteration $k$ is calculated as ${A_i^j}$/${{A_i}^k}$. It is determined by the area of the incremental region ${A_i^j}$ normalized with the unobserved area ${A_i}^k$.
\jw{Second, we measure the resolution of $\mathcal{R}_{j}$ as $G(\mathcal{R}_{j})$ by computing the average of the gradient magnitude over all pixels within the projection region~\cite{gal2010seamless}.} 
The photometric similarity, $C(\mathcal{R}_{j},\{ \mathcal{R} \})$, is defined by checking the photo-consistency of $\mathcal{R}_{j}$ against all the other projected images. It is computed by a multi-variate Gaussian function~\cite{waechter2014let}.
Furthermore, we introduce a new smoothness term to give a more confident value to a projected image $\mathcal{R}_j$ when $\mathcal{R}_j$ is consistent with the current set of selected images $\{\mathcal{I} \}_{i}^{k-1}$.
The final photometric quality term is organized as:
\begin{align}
&Q_{photo} =  [ \lambda_g G(\mathcal{R}_j) + \lambda_c C(\mathcal{R}_j,\{\mathcal{R}\}) ] \dfrac{A_i^j}{{A_i}^k} + \lambda_s Q_{smooth} ,\\
&Q_{smooth} = 1 - D_c (\mathcal{R}_j, \{\mathcal{I} \}_{i}^{k-1}) ,
\end{align}
where $\lambda_g$ and $\lambda_c$ are the respective weight of $G(\mathcal{R}_j)$ and $C(\mathcal{R}_j,\{ \mathcal{R} \})$, $\lambda_s$ is the weight of $Q_{smooth}$. \hq{$D_c (\mathcal{R}_j, \{\mathcal{I} \}_{i}^{k-1})$ represents the mean of pixel-wise color difference} between the overlapping pixels of image $\mathcal{R}_j$ and all the images in $\{ \mathcal{I} \}_{i}^{k-1}$ after normalization. $Q_{smooth}=1$ if iteration $k$ is zero or there exists no overlapping regions. The quality term $Q_{photo}$ tends to select a set of projected images with the maximum photometric and content consistency, which helps reducing the ghosting effect.
Note that both photometric and smoothness terms consider overlap regions. $C( \cdot )$ directly uses the whole projected regions (including overlap regions) of all the projected images $R$. Meanwhile, the smoothness term only considers the overlap region compared to the candidate projected image $I$.

\paragraph{Pespective quality.} However, only considering the photometric quality will possibly cause significant perspective inconsistency among selected regions/patches. An example is shown at the bottom left in Fig.~\ref{fig:texturing-issues}, where the partial region is from the left viewing angles and the other regions come from the right viewing angles. 
One solution is to evaluate one patch at a time and choose patches with viewing directions as close to the normal of the current proxy polygon as possible~\hq{\cite{wang2018seamless}}.
However, the set of candidate photos usually has a large variation in viewing direction and extremely uneven distribution, especially for those proxy polygons with large planes.
Strong front-parallel constraint could introduce missing regions while loose constraint could cause significant perspective differences.
To address this issue, we present a novel quality measurement to select a set of perspective consistent images, while satisfying the front-parallel constraint as much as possible.

The perspective quality term tends to select the next candidate image with a viewing direction similar to selected images in $\{ \mathcal{I} \}_{i}^{k-1}$, and with a viewing angle perpendicular to $\mathcal{P}_i$ as much as possible.
Hence, the perspective property is measured with front-parallel constraint and viewing angle consistency against selected photos. Unlike previous methods that only consider the viewing angle of a single candidate at a time~\hq{\cite{wang2018seamless}}, we propose to select a set of candidate images with high viewing angle consistency and as many front-parallel properties.
Photo collections usually have a large variety of viewing angles. In practice, it is difficult to find a set of $N$ photos from the input all with a high level of front-parallelism covering a large plane. Compared to focusing on front-parallelism, we hope to stress more on viewing direction consistency as $N$ gets larger. Hence, we introduce the inverse facade normal as the first selected viewing direction.
We define the perspective quality as:
\begin{equation}
Q_{persp}  = 1 - \dfrac{1}{N+1} [ D_{\hq{n}} (\mathbf{v}_j, -\mathbf{n}_i) + \sum_{n} D_{\theta} (\mathbf{v}_j, \mathbf{v}_n) ] ,
\end{equation}
where $N$ is the number of selected images in $\{ \mathcal{I} \}_{i}^{k-1}$, $\mathbf{n_i}$ is the plane normal of $\mathcal{P}_i$, $\mathbf{v}_j$ is the viewing direction of $R_j$, $\mathbf{v}_n$ is the viewing direction of the $n$-th image in $\{ \mathcal{I} \}_{i}^{k-1}$. \hq{$D_{n} (\mathbf{v},\mathbf{n}) = \frac{2}{\pi} \cos^{-1}{( \mathbf{v} \cdot \mathbf{n} )}$ calculates the normalized angular distance between facet normal $\mathbf{v}$ and plane normal $\mathbf{n}$.} $D_{\theta} (\mathbf{u}_1,\mathbf{u}_2) = \frac{\hq{1}}{\pi} \cos^{-1}{( \mathbf{u}_1 \cdot \mathbf{u}_2 )}$ calculates the normalized angular distance between facet normal $\mathbf{u}_1$ and $\mathbf{u}_2$.

\section{Line-guided Texture Mapping}
\label{sec:texture-mapping}

\begin{figure}[!t]
	\centering
	\includegraphics[width=.9\linewidth]{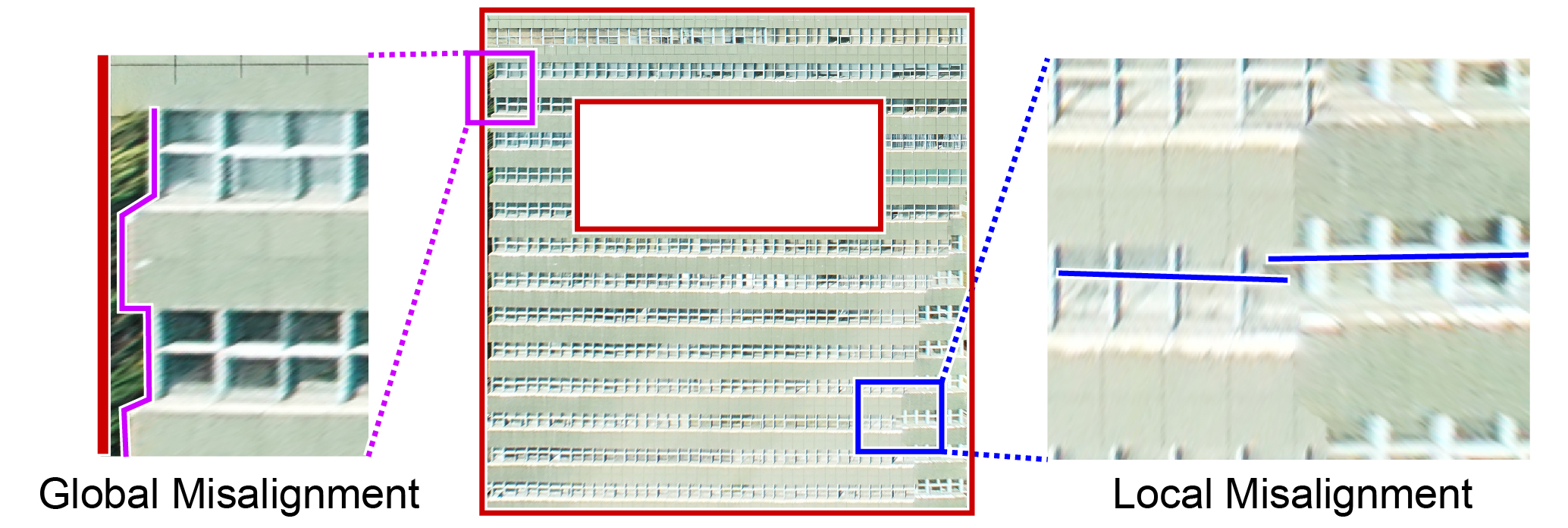}
	\caption{Simply stitching selected images and projecting to proxy polygon produces obvious geometric inconsistency. Left: global texture-to-geometry misalignment, where the magenta textural feature lines should match the proxy geometric boundary (visualized in dark red). Right: local texture-to-texture misalignment, where the blue feature lines should be connected and collinear.}
	\label{fig:misalignment}
\end{figure}

After determining a set of high-quality images ${ \{ \mathcal{I} \} }_{i}$ for each proxy polygon $\mathcal{P}_i$, our next goal is to generate a realistic and geometry-aware texture map for $\mathcal{P}_i$.
Urban buildings usually exhibit rich structural information, such as local/global line features (e.g., window border/windows layout) in facades, and structural line features of the overall building (e.g., facade border). Different from most existing texturing methods that are dealing with overall smooth models, the proxy model in our case is only $C0$ continuous with obvious sharp structural features. It makes our texture mapping a challenging problem. Even tiny texture-to-texture or texture-to-geometry misalignments will draw great visual attention, which can be observed in Fig.~\ref{fig:misalignment}.

Instead of relying on semantic label masks to synthesize textures~\cite {sinha2008interactive}, we make full use of the rich line features extracted from candidate views and the proxy boundary extracted from the planar polygon to guide the texture mapping.
Our algorithm begins by extracting three levels of line features (LoLs) from selected candidate images. A line-guided texture stitching approach is then proposed to improve the visual effects based on LoLs. Finally, we perform a texture optimization step to further improve the illumination consistency and texture completeness.
To further eliminate the seams across patches, we employ graph-cut image synthesis~\cite{kwatra2003graphcut, boykov2001fast} and Poisson-Image-Editing~\cite{perez2003poisson} to obtain one single enlarged texture\ignore{by stitching visible regions and optimizing the blended textures} for each proxy polygon.

\subsection{Levels of Line Features}
\label{sec:LOLs}

To generate proper line guidance at different scales we extract and optimize line segments from the level of local to global and finally define three levels of line features (LoLs). 

\paragraph{$LoL_0$.}
We first employ the ELSED algorithm~\cite{suarez2022elsed} to detect rich but discrete line segments from the projected images $\mathcal{R}$ as local features at the finest scale.
Then we organize these line segments into a set of global lines defined as $LoL_0$, in which each basic element encodes a meaningful primitive, \eg facade boundary, window boundary.
The $LoL_0$ is employed for rigid registration (the details will be described in Sec.~\ref{sec:texture-stitching}).
The registered images are then fed to later steps for generating $LoL_1$ and $LoL_2$.

\begin{figure}[!t]
	\centering
	\includegraphics[width=\linewidth]{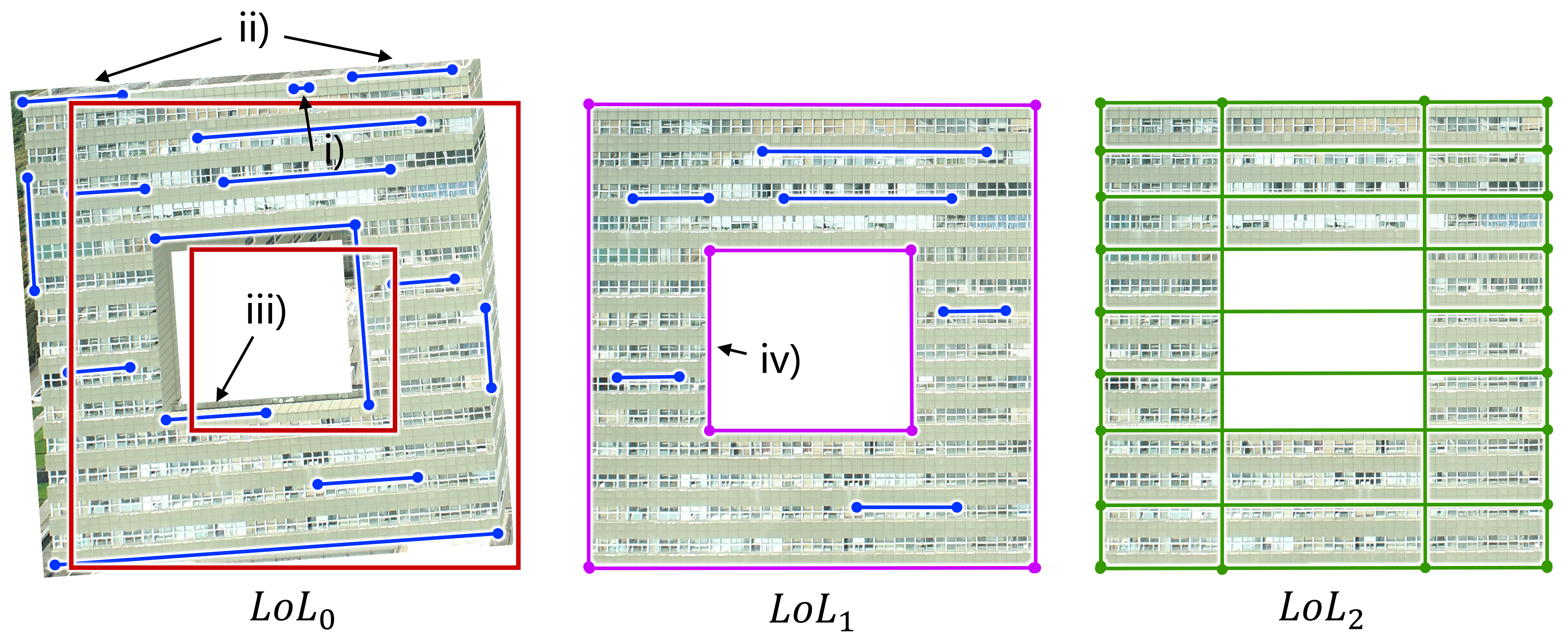}
	\caption{\wdx{Three levels of the line features extracted from a facade image from local to global (left to right). The respective proxy boundary is visualized in red. Some refinement operations i)-iv) on $LoL_1$ are pointed with arrows.}}
	\label{fig:LoLs}
\end{figure}

\paragraph{$LoL_1$.}
After registering image $\mathcal{I}_n$ via rigid alignment, we obtain its $LoL_1$ which has higher accuracy than $LoL_0$. However, the registered line segments still cannot well represent the polygon boundary. Our next step is to obtain $LoL_1$ which can maximally match $\mathcal{B}_i$.

We first measure the partial matching relationship of a line segment ${L} \in LoL_1$ compared to a boundary edge $b_m \in \mathcal{B}_i$ considering:
\begin{wrapfigure}{r}{0.35\linewidth}
	\centering
	\includegraphics[width=\linewidth]{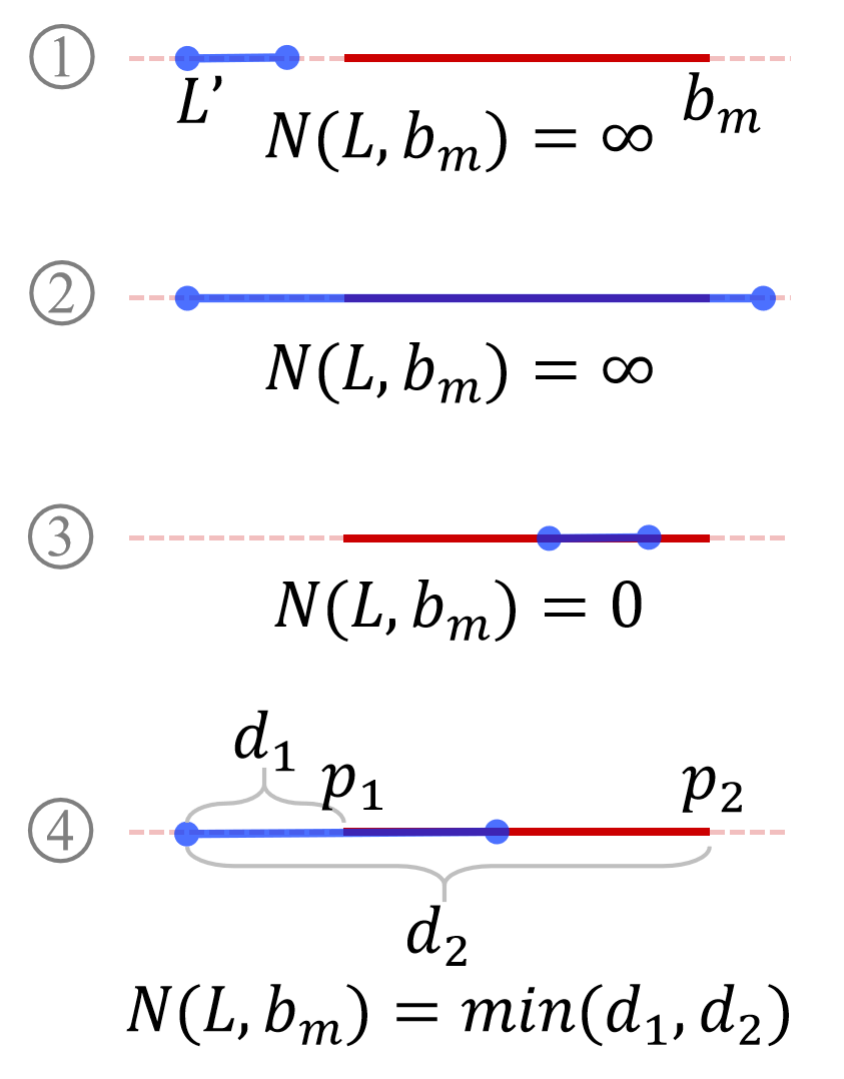}
\end{wrapfigure}
a) $\theta({L},b_m)$, the angle between two lines.
b) $D_{l}({L},b_m)$, the maximum point-to-line distance from end-points to another line.
Moreover, it is useful in matching line segments with similar slopes and spatial distance, but large variances in length.
c) $N({L},b_m)$, the level of non-overlap between ${L}$ and $b_m$. To compute it, ${L}$ is first projected to $b_m$ to get ${L}^{'}$.
We compute $N({L},b_m)$ based on four types of relationship between ${L}^{'}$ and $b_m$ after line projection, see the inset figure.
In case 1 and 2, we assign infinity to $N({L},b_m)$. In case 3, $N({L},b_m)$ is set to zero. A point projected on $b_m$ with end-points $p_1$ and $p_2$ can be denoted as $p_1 + w(p_2-p_1)$. In the last case, for the end-point of ${L}^{'}$ locating outside $b_m$ ($w<0$ or $w>1$), we calculate its distance to the end-points of $b_m$ and assign the minimum value to $N({L},b_m)$.
${L}$ with value of a)-c) smaller than given thresholds will be denoted as a boundary matching line $\hat{L}$.

The set of lines of $LoL_1$ are further refined by: i) eliminating tiny noisy $\hat{L}$ matching any $b_m$, ii) merging the set of lines $\hat{L}$ matching the same $b_m$ into one single line segment, iii) extending $\hat{L_1}$ and $\hat{L_2}$ until they share a same end-point, if the matching boundary $b_p$ and $b_q$ of $\hat{L_1}$ and $\hat{L_2}$ are neighbors, and iv) connecting $\hat{L_1}$ and $\hat{L_2}$ whose matching boundary are $b_p$ and $b_q$ respectively, if the inserted $L$ can match $b_o$ whose neighbors are $b_p$ and $b_q$.
The above operations are illustrated in Fig.~\ref{fig:LoLs} and the refined $\hat{L}$ are visualized in purple.
The $LoL_1$ can serve as a guidance for texture stitching in Sec.~\ref{sec:texture-stitching}.

\paragraph{$LoL_2$.}
Finally, we cluster all the line segments in $LoL_1$ to extract structural lines with a dynamic K-means algorithm, according to the slope and the distance from the image center to the line segment.
After that, we can obtain $LoL_2$ to represent the structural layouts of facade components and to serve as a reference for later inpainting in Sec.~\ref{sec:texture-optimization}. Until now, for each selected image, we obtain LoLs for different operators. Fig.~\ref{fig:LoLs} illustrates the LoLs from a facade image of Hitech.

\subsection{Line-guided Texture Stitching}
\label{sec:texture-stitching}

For each proxy polygon $\mathcal{P}_i$, 
we have obtained a subset of views $\{ \mathcal{I} \}_i$ along with their LoLs in previous steps.
For simplicity, we denote the first selected image $\mathcal{I}_1$ in $\{ \mathcal{I} \}_i$ as a reference image $\mathcal{I}^{ref}$ and the others as target images $\mathcal{I}^{tar}$.
To generate an enlarged texture map, we need to find the matching relationship between two images to stitch them together.
Based on extracted LoLs, one solution is to find the correspondence between the lines (\eg using LBD descriptor~\cite{zhang2013efficient}) by considering both image and geometry features.
However, there exists plenty of repetitive patterns which make such pixel-based methods less effective. Therefore, we develop an algorithm for matching line segments by utilizing geometric information.

Due to the camera and geometry inaccuracy, there exists obvious texture-to-geometry displacement which increases the difficulty in accurate image local matching and stitching. We start by globally registering $\mathcal{I}^{ref}$ towards $\mathcal{B}_i$ with a rigid transformation and several warping operators. Next, following the order of their insertion into $\{ \mathcal{I} \}_i$, all the $\mathcal{I}^{tar}$ are sequentially deformed to match $\mathcal{B}_i$ globally, then locally stitched to $\mathcal{I}^{ref}$. The $n$-th inserted image is denoted as $\mathcal{I}_{n}$.
The stitching process is done in three steps: rigid alignment, line matching and texture warping.

\paragraph{Rigid alignment.}
To reduce the large texture-to-geometry misalignment, each image $\mathcal{I}_{n}$ is transformed toward the proxy boundary $\mathcal{B}_i$. Firstly, we extend each image with a margin of $N$ pixels and extract complete $LoL_0$ from the extended version. We compute the bounding box of $\mathcal{B}_i$ and set $N$ to $5\%$ of the length of its diagonal. This can be observed in Fig.~\ref{fig:LoLs}.
Next, we convert all the endpoints of $LoL_0$ in $\mathcal{I}_n$ into a source point cloud $\mathcal{C}_n$. We then convert the endpoints of $\mathcal{B}_i$ into a target point cloud $\mathcal{C}_i$. A rigid transformation is calculated via ICP~\cite{besl1992method} to align $\mathcal{C}_n$ toward $\mathcal{C}_i$. This rigid transformation is employed to transform and update $\mathcal{I}_n$.

\paragraph{Line segment matching.} We denote $LoL_1$ of $\mathcal{I}^{tar}$ as $\mathcal{ {L}}^{tar}$ and $\mathcal{I}^{ref}$ as $\mathcal{ {L}}^{ref}$. For each line segment $ {L}^{tar} \in \mathcal{ {L}}^{tar}$, it can be registered to line segment $L^{ref} \in \mathcal{L}^{ref}$ if satisfying: i) the angular distance $\theta({L}^{tar},L^{ref})$ is smaller than $5^{\circ}$, ii) the point-to-line distance $D_{l}({L}^{tar},L^{ref})$ is closer than 10 pixels, and iii) the level of non-overlap $N({L}^{tar},L^{ref})$ is smaller than 100 pixels if $L^{ref} \in \mathcal{B}_i$. We denote $\mathcal{L}$ as the set of all pairs of matched line segments between $ {\mathcal{L}}^{tar}$ and $\mathcal{L}^{ref}$.

\paragraph{Texture warping.} With the matched lines in $LoL_1$, we can warp $\mathcal{I}^{tar}$ to $\mathcal{I}^{ref}$ while preserving the matched line segments. The warping of $\mathcal{I}_n$ to $\mathcal{B}_i$ shares the same process\ignore{ with matched lines in $LoL_1$}.
Previous work~\cite{jia2021leveraging} converts the transformation into a global line-guided mesh deformation problem. They rely on deforming the vertices of a grid-mesh uniformly sampled in the target image to maintain line structures by solving a global least-square energy function.

However, the global optimal deformation method may not guarantee the strict straightness of colinear line features in facade images.
\begin{wrapfigure}{r}{0.25\linewidth}
	\centering
	\includegraphics[width=\linewidth]{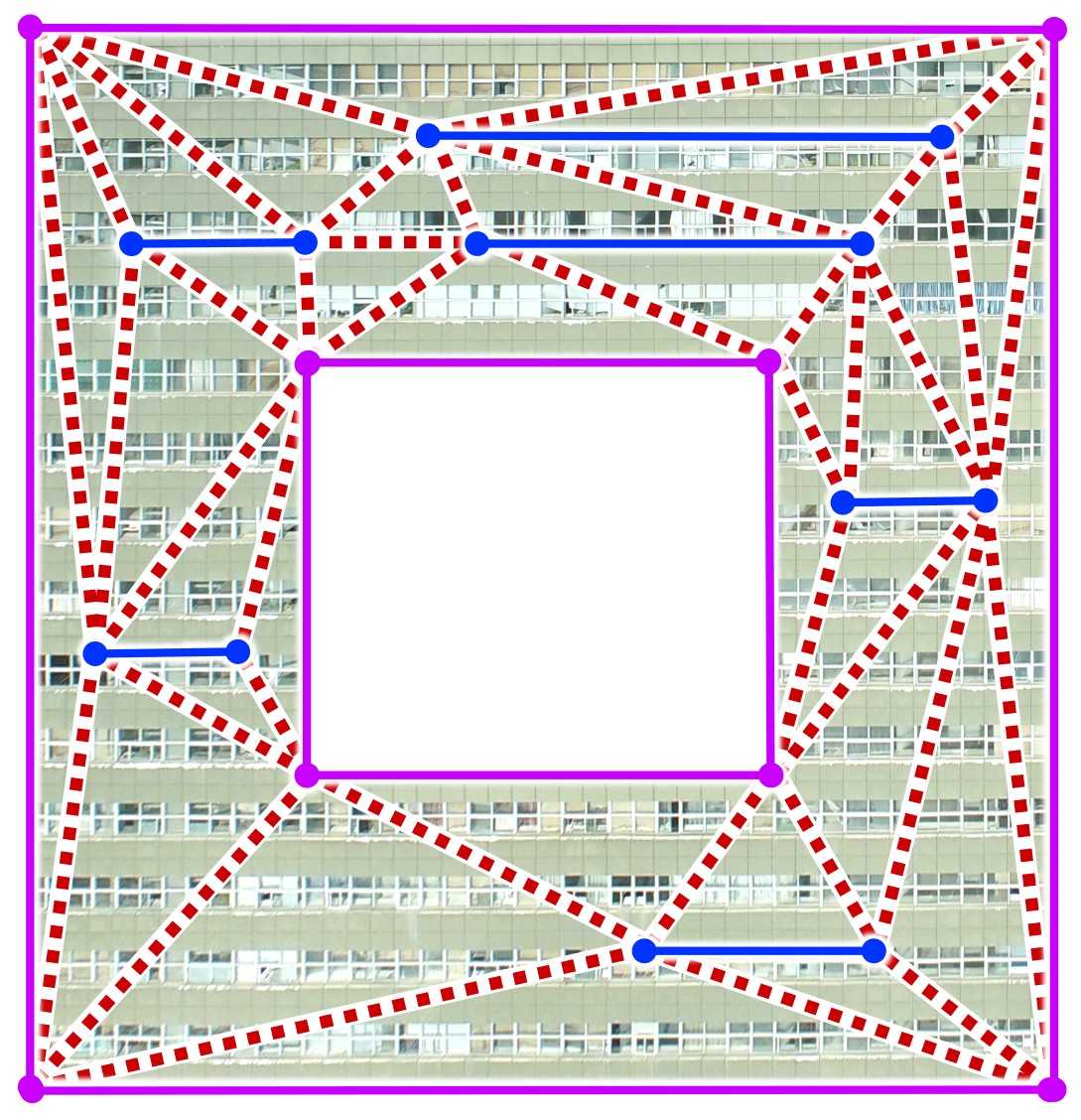}
\end{wrapfigure}
In this work, we employ an adaptive-mesh data structure to explicitly maintain the line segments after warping. First, we construct a constrained Delaunay triangulation $M^{tar} = (V^{tar},E^{tar},F^{tar})$ based on $LoL_1$, keeping all the inserted segments as constraint edges, as shown in the right wrapped figure. Unlike uniformly sampling grid-mesh data structure, every basic element in our adaptive mesh encodes a meaningful geometric primitive in the target image, \ie a corner point or a line segment.
Based on this, all the vertices $V^{tar}$ are deformed to new positions $\widetilde{V}^{tar}$ while satisfying the following two constraints: 
\begin{itemize}[leftmargin=*]
	\item The matched line segment pairs in $\mathcal{L}$ are well aligned after the mesh deformation.
	\item All the line segments in $ {\mathcal{L}}^{tar}$ preserve straightness after the mesh deformation.
\end{itemize}

We search for the optimal deformation offset $\Delta{V}=\widetilde{V}^{tar}-V^{tar}$ by embedding these two constraints and a normalization term into an energy minimization formulation:
\begin{equation}
	E(\Delta{V})=\lambda_a E_a(\Delta{V}) + \lambda_l E_l(\Delta{V})   + \lambda_r E_r(\Delta{V})\text{.}
 \label{eq:deformation}
\end{equation}

$E_a(\Delta{V})$ represents the line-alignment term, which guarantees the alignment between all pairs of matched line segments in $\mathcal{L}$ after deforming vertices by $\Delta{V}$ in the form of
\begin{equation}
	E_a(\Delta{V}) = \sum_{t=1}^{|\mathcal{L}|}d(V^{tar}_{t1}+\Delta{V_{t1}}, l_t^{ref})+ d(V^{tar}_{t2}+\Delta{V_{t2}}, l_t^{ref}) \text{,}
\end{equation}
where $l_t^{ref}$ is the $t$-th matched line segment in $\mathcal{L}^{ref}$, and $(V^{tar}_{t1},V^{tar}_{t2})$ are end points of $t$-th matched line segment in $ {\mathcal{L}}^{tar}$. Moreover, the line-preserving term denoted as $E_l(\Delta{V})$ ensures the straightness of each line segment in $ {\mathcal{L}}^{tar}$ after mesh deformation by
\begin{equation}
	E_l(\Delta{V}) =\sum_{j=1}^{| {\mathcal{L}}^{tar}|} [ (V^{tar}_{j1}+\Delta{V_{j1})-(V^{tar}_{j2}+\Delta{V_{j2})} ]^T \cdot \overrightarrow{n}_j^{tar}} \text{,}
\end{equation}
where $(V^{tar}_{j1},V^{tar}_{j2})$ are end points of $j$-th line segment $l_j^{tar}$ in $ {\mathcal{L}}^{tar}$ and $\overrightarrow{n}_j^{tar}$ is the normal vector of $l_j^{tar}$.
The normalization term to prevent exaggerate deformation is defined as $E_r(\Delta{V}) = \sum_{t=1}^{|\mathcal{L}|} ||\Delta{V}||^2$. $\lambda_a$, $\lambda_l$, and $\lambda_r$ in Eq.~\ref{eq:deformation} is set to 0.5, 0.5, and 0.025 in all experiments.

Instead of optimizing the vertex position via LSQM~\cite{jia2021leveraging}, we calculate the vertex offset by applying the conjugate gradient algorithm which is more stable and efficient.
Unlike using the calculated vertex position to update enclosing grid vertices with bilinear interpolation~\cite{jia2021leveraging}, each line segment represents a linear feature of a facade component in our case. We directly apply the optimal vertex offsets to the end-points of lines for the preservation of linear features.
There can appear topological issues of the updated mesh after that. Hence, we introduce a refinement step to address the topological issues such as line crossing and non-manifolds.
Finally, with an affine transformation constructed by the refined vertices of adaptive mesh facet $f \in F^{tar}$, each pixel $p^{tar}$ inside the $f$ is warped to the new position $\widetilde{p}^{tar}$.
With the alignment and warping operators, all the visible regions are globally warped to $\mathcal{B}_i$ and locally warped to $\mathcal{I}^{ref}$ forming an enlarged texture image for each proxy polygon $\mathcal{P}_i$.

\subsection{Texture Optimization}
\label{sec:texture-optimization}

\paragraph{Illumination adjustment.} Previous steps enable us to generate a texture map $\mathcal{I}_i$ for each proxy polygon $\mathcal{P}_i$. Since $\mathcal{I}_i$ is composed of patches extracted from various viewing directions, the brightness may be inconsistent across neighbouring patches. Assuming that the brightness distribution over the whole texture map should hold a specific pattern, we perform a histogram specification operation to improve the illumination consistency of $\mathcal{I}_i$ in HSV color space. To be precise, we first calculate a distribution histogram of the channel $V$ on the non-overlapping region of $\mathcal{I}^{tar}$ and overlapping region of $\mathcal{I}^{ref}$, respectively. Next, we match these two histograms and transfer the brightness of the overlapping region on $\mathcal{I}^{ref}$ to the non-overlapping region on $\mathcal{I}^{tar}$.

\paragraph{Texture inpainting.} In practice, some regions in $\mathcal{P}_i$ may not be observed by any of the selected views or covered with high-quality patches, leading to holes in the generated texture map. To reduce this visual artifacts, we need to fill the missing regions (referred to as masks) to generate a complete texture map. However, the arbitrary sizes and irregular shapes of the mask pose great challenges to previous inpainting methods (\eg patch-based approaches~\cite{barnes2009patchmatch,huang2014image}). Inspired by the strong capability of diffusion model in generating semantic and geometric harmonious results, we employ a denoising diffusion probabilistic model called RePaint~\cite{lugmayr2022repaint} for free-form inpainting.

\begin{figure}
	\centering
	\includegraphics[width=\linewidth]{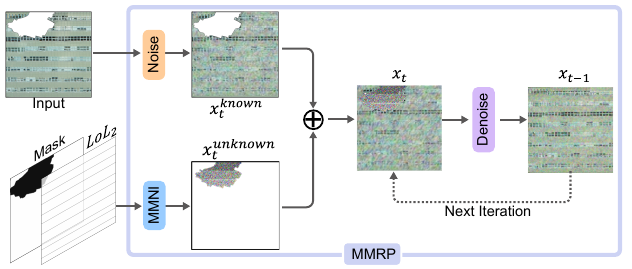}
	\caption{Overview of our proposed texture inpainting approach, MMRP.}
	\label{fig:inpainting-overview}
\end{figure}

However, RePaint is still not robust enough to generate structurally consistent results for a large missing region. To alleviate this issue, we further introduce a line-guided version of RePaint called Multi-Mask RePaint model (MMRP) to preserve the geometric consistency.
As shown in Fig.~\ref{fig:inpainting-overview}, the input of MMRP contains the mask, $LoL_2$, and the texture to be completed.
We then propose a multi-mask noise initialization (MMNI) component.
Firstly, we compute the principal directions of the lines in $LoL_2$ by using principal component analysis (PCA). Our MMNI divides the unobserved region into several parts following the second max principal direction.
As mentioned by~\citet{croitoru2023diffusion}, the results will show better Frechet Inception Distance values and faster convergence with a mixture of Gaussian noises. Inspired by this observation, unlike the original RePaint that initializes the entire unobserved region with a single random Gaussian noise, we randomly initialize each divided part with a random Gaussian noise $N(\mu,\sigma)$, where $\mu$ ranges from [0, 10], $\sigma$ ranges from [1, 50]. The ranges are chosen from experimental results that deliver good performance.

After that, the parts initialized with multiple noises along with observed content are merged into one image, $x_{t}$, which will be harmonized together into a new inpainting.
With MMNI, our MMRP can increase the possibility of generating result that is more consistent with the observed region.

We find that the quality of training data is crucial for diffusion model to generate feasible results for a specific scenario, such as the architecture in our case. To adapt RePaint to our task, we create a TwinTex dataset and fine-tune Repaint which was originally trained on ImageNet~\cite{russakovsky2015imagenet}. 
Starting from $44k$ high-resolution images collected with aerial drone covering $17$ outdoor architectural scenes, we first project each image to all its visible $\mathcal{P}_i$ and extract the pixels inside the bounding box of visible regions. Each extracted image is then cropped to a set of images with the size of $512 \times 512$. Next, we remove the images with heavy blurring issues or containing a large portion of non-architectural contents (\eg vegetation, human, scaffolds, etc)~\cite{zhao2017pyramid}. Our final TwinTex dataset (TwinTexSet) contains $46k$ images of various scene components and building categories (school, office building, etc). Note that all the image regions with heavy blurring issue and extreme inclined angles have been removed from our data set.
Note that there are some overlaps between the training and testing scenes. However, the TwinTex dataset contains the processed photos of 17 outdoor scenes, where we have rectified the photos, removed blurring regions and large non-architectural objects. After these processes, the training images and the test texture maps have no overlap. The missing regions in test images come from frustum-culling, visibility detection, or are not included in input photos. Therefore, the regions to be infilled never appear in the training images, thus never seen by the trained model in our tasks.

We make use of our TwinTex dataset to fine-tune the pre-trained RePaint model. The total training time is about 9 hours.  We only need to pay the time cost for training once.
For an image with high resolution, we will resize it to $512 \times 512$. 
And for an image with large missing area, we choose to scale it according to the largest dimension while preserving the image ratio. The resized image will be fed into MMRP to receive a semantic and geometric consistent inpainting. Finally, we upsample the resultant inpainting to the original size with a bi-linear interpolation operation.
Please refer to the supplemental for more details and experimental results.

\section{Results}
\label{sec:results}

To validate the proposed algorithm, we conduct a series of experiments on real-world scenes with varying building styles and functions.
We first demonstrate the robustness of our approach and the effectiveness of each main ingredient through a stress test.
Then, we evaluate our approach by comparing it against state-of-the-art texture mapping methods.
Ablation studies are performed as well to show the effectiveness of our main technical ingredients (please refer to the supplemental material for the details).
Texturing results by our method and manual work on proxy models at different levels of detail are also provided in the supplemental material to demonstrate the high quality of our generated texture maps. 
Our algorithm is implemented in C++ and also well-organized as a customized plug-in for Houdini. All the presented experimental results are obtained on a desktop computer equipped with an Intel i9-10900k processor with 3.0 GHz and 128 GB RAM. Note that the time cost for fine-tuning the RePaint model are not included in the statistics.

\paragraph{Dataset.}
For the performance evaluation and comparisons, we carry out experiments on \wdx{$18$ real-world scenes, including $15$ outdoor buildings, two indoor rooms and a man-made object}. These scenes come from public sources (\ie Library~\cite{DronePath21}, Hitech~\cite{DroneScan20}, Polytech and ArtSci~\cite{UrbanScene3D}), a handheld device (\ie Machine Room, Lab, Cabinet), or are captured by a single-camera drone (\ie Highrise, School, Center, CSSE, Sunshine Plaza, Hall, Center, Factory, Hisense, Bank, Apartment).
Please refer to the supplemental material for detailed statistics on the photo collections and models.
Then we use the commercial software RealityCapture\footnote{https://www.capturingreality.com/} (RC) to reconstruct the 3D high-precision models from the captured images.
All the abstracted versions are either the 2.5D models for path planning~\cite{DroneScan20}, or created by in-field modelers.
All of the dense reconstructions and proxy models are utilized to demonstrate the performance of our algorithm.

\subsection{Stress Test}

\begin{figure}
	\centering
	\includegraphics[width=.96\linewidth]{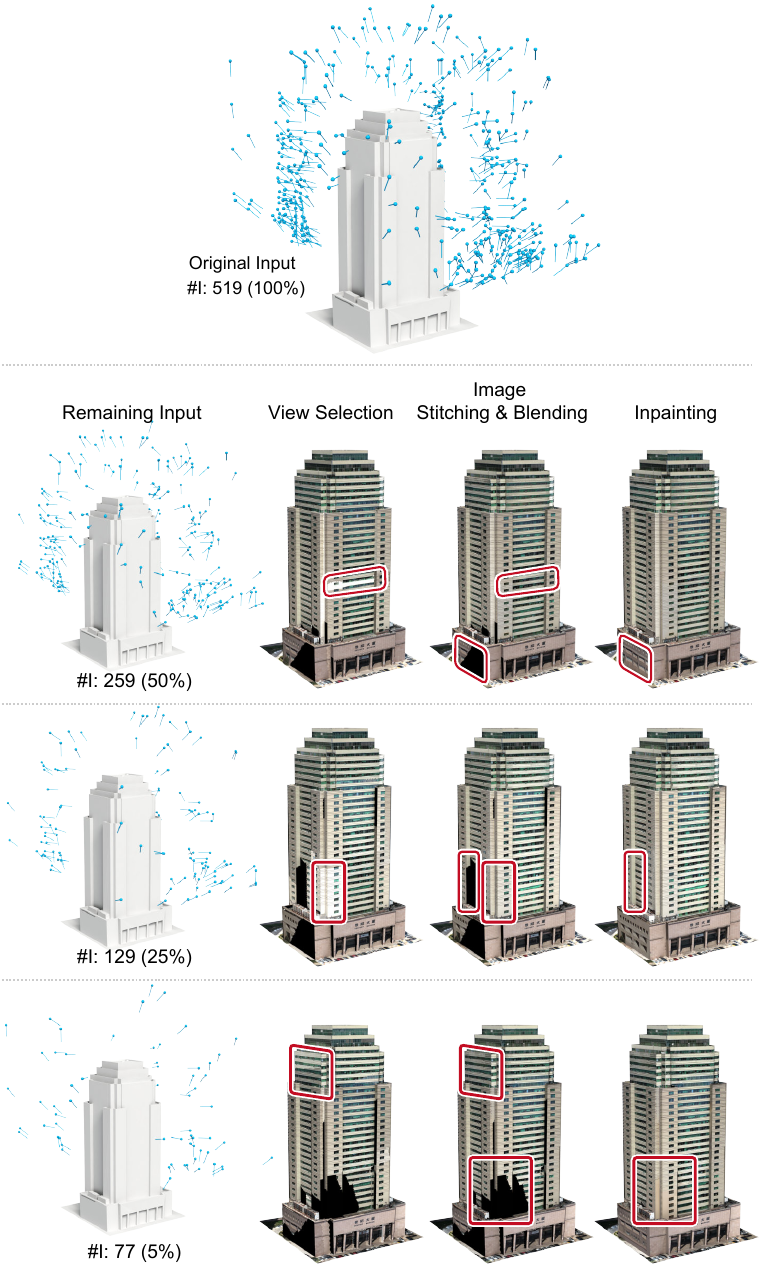}
	\caption{Stress test of our methodology on Hall example via randomly removing a certain percentage of photos from original input. In each row we visualize: the camera distribution of the remaining input, the textured proxy based in remaining input after view selection, image stitching with blending, and inpainting. $\# I$ denotes the number of photos. The value in the bracket denotes the percentage of the remaining photos related to original input.}
	\label{fig:stress-test}
\end{figure}

We first evaluate the robustness of our approach on a complicated Hall example with 519 input views. This stress test is performed via randomly removing partial photos from the original photo set and feeding the rest views into our TwinTex. The textured proxies of Hall example generated with three different sets of remaining views are shown in Fig.~\ref{fig:stress-test}. We also show the intermediate texture maps after each step, \eg view selection, image stitching along with blending and texture inpainting.

\begin{figure*}[tbh!]
	\centering
	\includegraphics[width=\linewidth]{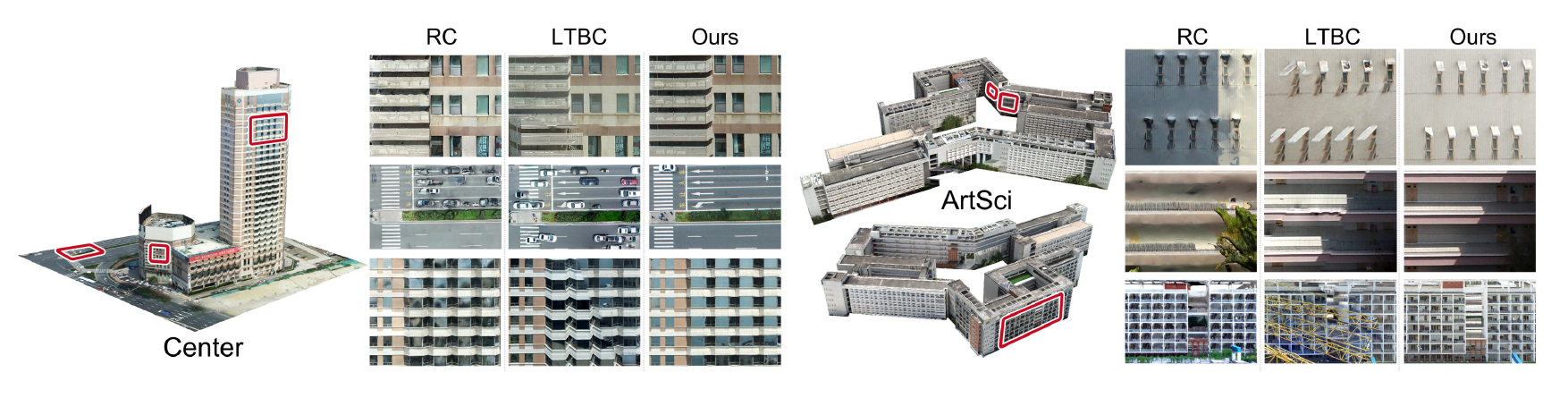}
	\caption{Visual comparison against RC, LTBC and ours on two examples. The detailed comparisons are shown in the zoomed-in insets. Due to the page limit, we only show the overall views of our textured proxy models, which are shown on the left to the zoomed-in insets.}
	\label{fig:comparison}
\end{figure*}

We randomly removed $50\%$, $75\%$ and $95\%$ photos from the original set and made use of the remaining views to texture this proxy model.
The textured proxies after view selection are generated by simply overlapping the projection of selected images following the reverse order of selection. As the number of photos in the remaining input decreased, our selection algorithm can still select high-quality photos from the input to maximally cover the proxy, as illustrated in the second column of Fig.~\ref{fig:stress-test}. From the third column, we can see the texture maps after performing our stitching step and illumination adjustment exhibit higher geometric and photometric consistency. However, there appears larger missing regions as the number of remaining views decrease.
We only use $77$ images to texture Hall example with $225$ facades (each plane has about $6$ photos to select from).
All the missing regions are visualized with black pixels, as marked by the red boxes in the figure. The final texture maps visualized in the last column of Fig.~\ref{fig:stress-test} demonstrate the ability of our inpainting framework for creating geometrically and semantically consistent contents under very limited resource.

\subsection{Comparison with Texturing Methods}

In this subsection, we evaluate the quality of our generated texture maps by comparing them against RC and two competitive texturing generation frameworks, let there be color (LTBC)~\cite{waechter2014let} and patch based optimization (PBO)~\cite{bi2017patch}.
We adopt the evaluation scheme proposed by~\citet{Waechter2017} to compare the rendered image with the corresponding real image following a specific view from the input cameras.
We select the ground truth photos for evaluation from the input excluding the photos for texturing.
Two visual similarity metrics, namely the structural similarity index measure (SSIM) and learned perceptual image patch similarity (LPIPS)~\cite{kastryulin2022piq}, are adopted for quantitative evaluation.

\textbf{Comparison with LTBC.}
We first compare our method with RC and LTBC on two examples.
Note that LTBC selects the most suitable view for each facet by solving a Markov Random Field formulation and adjusting the pixel values along the boundary edges of adjacent patches to alleviate the seams. This strategy fails to texture large facets when there is no camera view observing the whole region of the facet. Thus, we subdivide the simplified coarse model and perform LTBC to generate texture maps for the corresponding subdivided mesh for a reasonable comparison.

\begin{table}
	\caption{Quantitative comparison on the texturing results in Fig.~\ref{fig:comparison}.}
	\label{tab:quality-results}
	\begin{center}
		\begin{tabular}{c c c c c}
			\midrule[1pt]
			\multirow{1}{*}{Scene} & \multirow{1}{*}{Method}  & $ SSIM\uparrow$ & $LPIPS \downarrow$   & Time (min)\\
			\midrule
			\multirow{3}{*}{Center}   & RC      &    0.350     &  0.831    & 287.0    \\
			\multicolumn{1}{c}{}        & LTBC  &    0.354  &   0.836    & 4.4   \\
			\multicolumn{1}{c}{}        &  Ours     & \textbf{0.358}       &   \textbf{0.831}  &  167.4   \\ \hline
			\multirow{3}{*}{Artsci}   & RC     &   0.346    & 0.883   &   320.9   \\
			\multicolumn{1}{c}{}        & LTBC      &  0.331   &      \textbf{0.873}    &   6.0  \\
			\multicolumn{1}{c}{}        &  Ours                        & \textbf{0.348}   &      0.877    &     256.1  \\
			\midrule[1pt]
		\end{tabular}
	\end{center}
\end{table}

\begin{table}
	\caption{Quantitative comparison on the performance of texturing results on the third and fourth examples in Fig.~\ref{fig:comparison2}. Numbers in the bracket denote the quality value of the three zoomed-in views of each example from top to bottom. All the times are in minutes.}
	\label{tab:quality-results2}
	\begin{center}
		\begin{tabular}{c c c c c}
			\midrule[1pt]
			Scene & Method  & $ SSIM\uparrow$ & $LPIPS \downarrow$    & Time\\
			\midrule	  									
			\multirow{4}{*}{\small{Library}}   & \small{RC}      &  \small{\{ 0.28, 0.28, 0.32\ignore{,0.53} \}}   &  \small{\{ 0.53, 0.60, 0.68\ignore{,0.65} \}}       & \small{383.6}       \\
			& \small{LTBC}   &  \small{\{ 0.29, 0.29, 0.34\ignore{,0.46} \}}   &  \small{\{ 0.54, 0.68, 0.68\ignore{,0.48} \} }      & \small{4.8}         \\
			\multicolumn{1}{c}{}        & \small{PBO}   &  \small{\{ \textbf{0.63}, \textbf{0.45}, 0.38\ignore{,\textbf{0.74}} \}}  &  \small{\{ 0.63, 0.67, 0.69\ignore{,0.84} \}}     & \small{810.7}        \\
			&  \small{Ours}     &  \small{\{ 0.46, 0.38, \textbf{0.36}\ignore{,0.46} \}}   &  \small{\{ \textbf{0.28}, \textbf{0.52}, \textbf{0.66}\ignore{,\textbf{0.47}} \}}     & \small{200.4}             \\
			\hline		
			\multirow{4}{*}{\small{Bank}}   & \small{RC}      & \small{\{ \ignore{0.61,}0.25, 0.62, 0.26 \}} & \small{\{ \ignore{0.28,}0.43, 0.23, 0.54 \}} &      \small{198.3}         \\
			& \small{LTBC}   & \small{\{ \ignore{0.67,}0.20, 0.52, \textbf{0.33} \}} & \small{\{ \ignore{0.19,}0.51, 0.37, 0.53 \}}       & \small{6.4}   \\			
			& \small{PBO}   & \small{\{ \ignore{0.65,}0.25, \textbf{0.70}, 0.27 \}}   & \small{\{ \ignore{0.22,}0.43, 0.31, 0.57 \}}       &   \small{672.1}      \\
			&  \small{Ours}  &  \small{\{ \ignore{\textbf{0.97},}\textbf{0.26}, 0.63, 0.28 \}}   &  \small{\{ \ignore{\textbf{0.02},}\textbf{0.27}, \textbf{0.18}, \textbf{0.52} \}}    &  \small{174.5}          \\
			\midrule[1pt]
		\end{tabular}
	\end{center}
\end{table}%

\begin{figure*}[tbh!]
	\centering
	\includegraphics[width=.76\linewidth]{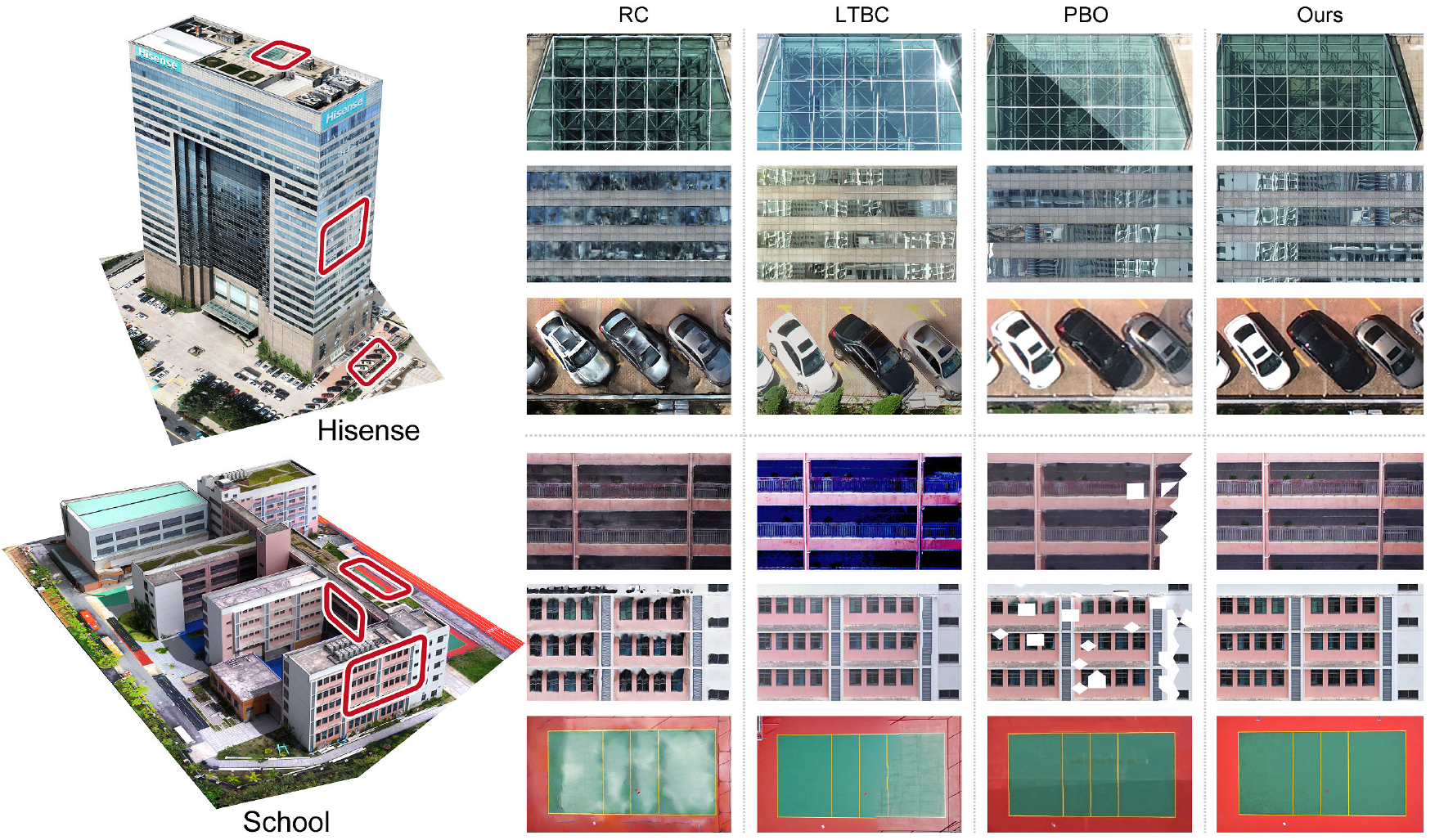}\\
	\includegraphics[width=.76\linewidth]{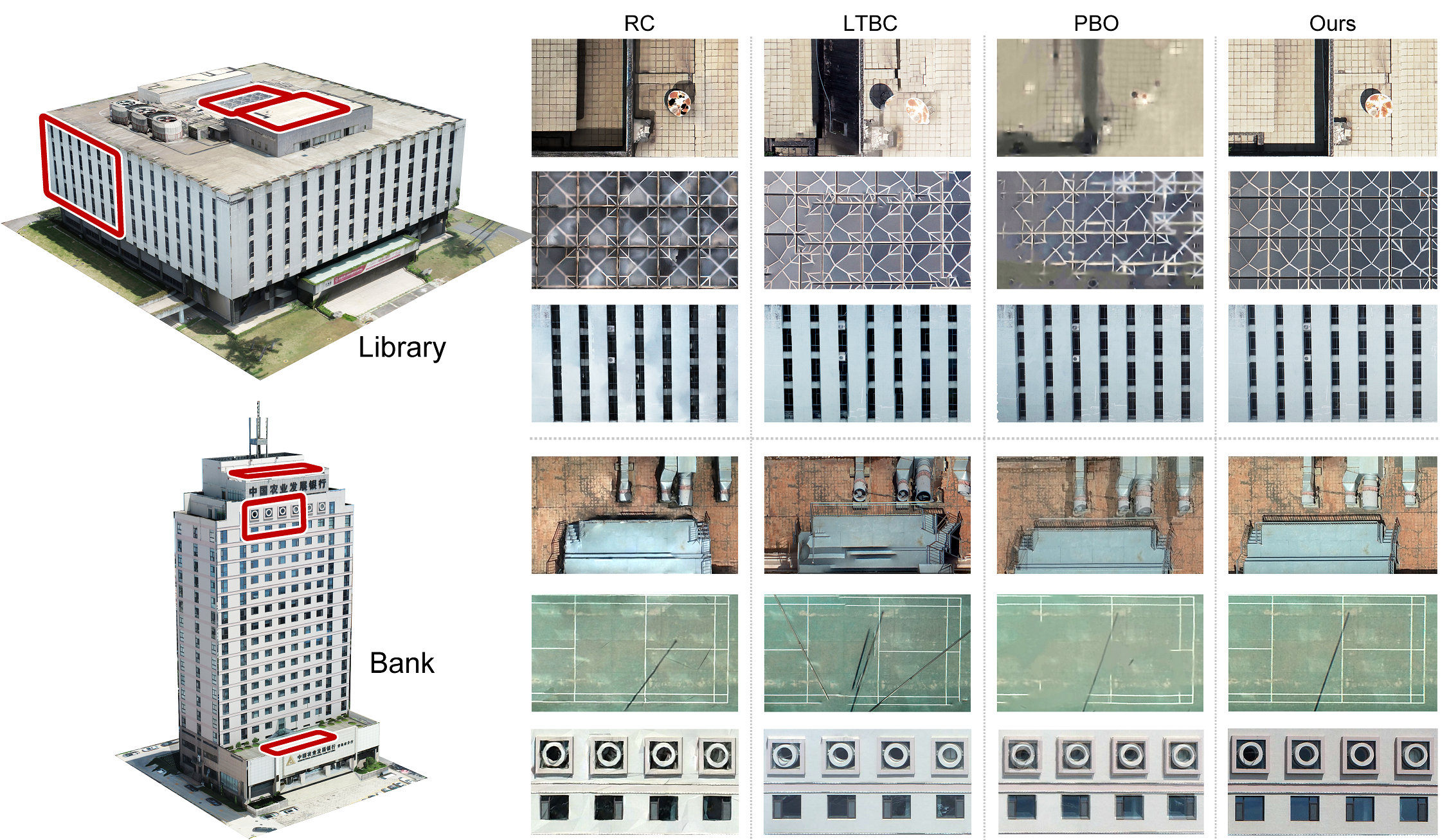}
	\caption{Visual comparison against RC, LTBC, PBO and ours on four examples. The detailed comparisons are shown in the zoomed-in insets.}
	\label{fig:comparison2}
\end{figure*}

\begin{figure*}[!tbh]
	\centering
	\includegraphics[width=\linewidth]{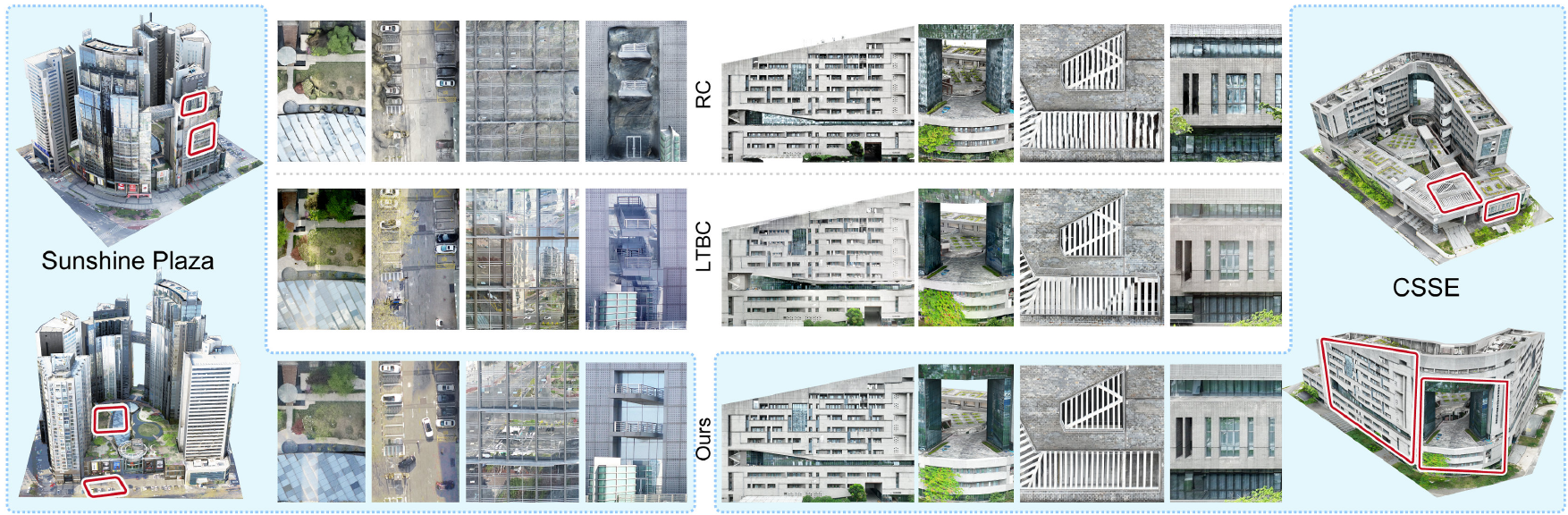}
	\caption{Visual comparison of two challenging cases against RC, LTBC and ours. The detailed comparisons are shown as zoomed-in insets.}
	\label{fig:challenging-cases}
\end{figure*}

Fig.~\ref{fig:comparison} shows the qualitative results on two abstracted models with sharp textures and structural features.
First, we can observe that LTBC and RC are not able to produce satisfactory results for large-scale planar structures which are invisible for the input photos or with only low-quality pixels. In these cases, their texture maps show blurring and ghosting artifacts. By contrast, our method can fill large unobserved regions with geometrically and semantically consistent pixels.

Second, without the constraints on perspective consistency in view selection, LTBC failed in selecting photos with both consistent and front-parallel viewing directions, creating "stretching" alike artifact and perspective inconsistency in the generated textures. The "stretching" problem can be observed in the inset of the ArtSci building in Fig.~\ref{fig:comparison} (first row, second column), which was caused by merging photos with large viewing angle variance. The perspective inconsistency can be observed in the insets of the Center example in Fig.~\ref{fig:comparison} (first and third rows, second column). Unlike LTBC, our view selection strategy can pick out a very small set of perspective consistent and front-parallel photos as possible. Moreover, we introduced a smooth term considering the pixel-wise distance of each image pair. This helps select a set of photos with a higher level of photometric and content consistency, especially for filtering out occluders or dynamic instances (e.g., cars). The significant improvement can be observed in the Center example in Fig.~\ref{fig:comparison} (second row) and ArtSci example in Fig.~\ref{fig:comparison} (third row).

Furthermore, due to the lack of explicit constraints on the linear features of buildings, the texture generated by LTBC destroys geometric structures, making them suffer from serious seams and distortion effects.
Our image stitching mechanism succeeds in preserving the straightness and completeness of the line segments by constructing an adaptive mesh taking the original geometric primitives as constraints.
Next, large and duplicate texture patterns are also well maintained in our results. This is mainly because we select a suitable small subset of camera views for each planar shape instead of each facet. This strategy guarantees that each selected view is used to texture the current plane as much as possible.

The quantitative comparison results are given in Table~\ref{tab:quality-results}. In most of the cases, the rendered views of our textured model can deliver better SSIM and LPIPS scores. Please refer to the supplemental material for the overall views of the above two examples.

\textbf{Comparison with PBO.}
Next, we perform comparison with PBO\footnote{https://github.com/yanqingan/EAGLE-TextureMapping} and report SSIM and LPIPS as well on four more architectural scenes.
We first tried to globally compare the similarity of rendering views on the entire building with real photos. However, PBO requires an extremely large amount of time/storage given just several high-resolution images as input. Remember that our target is to retain the quality of the inputs, one way to address the issues is to decrease the number of input photos. This can produce other problems such as missing contents since no pixels can cover some of the regions. It is crucial to provide key views with high quality to the PBO in this scenario.
Hence, we locally perform the comparison to PBO using three planes per example. To this end, we fed PBO with the set of views selected by our algorithm for the planar polygons involved in the comparison.
Considering the nature of patch-based optimization, for a fair qualitative comparison, we use the subdivided mesh prepared for LTBC as the input to PBO as well.
Fig.~\ref{fig:comparison2} shows the texturing results of RC, LTBC, PBO and ours. The comparison shows that given the same set of selected views, PBO can generate a texture with good quality in most cases which also proves the effectiveness of our view selection algorithm. However, the results of PBO have obvious blurring issue and line distortion artifact compared to the real photos. There are also some optimization failures of PBO which cause missing textures for the subdivided triangles. In comparison, our stitching and optimization frameworks can preserve clear line features and retain the original structure while we are able to inpaint missing regions.

Quantitative results are listed in Table~\ref{tab:quality-results2}.
Since SSIM is by nature less sensitive to blurring issues and possibly gives higher scores to images with such artifacts~\cite{zhang2018unreasonable}, the texture with the regional blurring problem could have higher scores. This phenomenon can be observed in some views per each example. Meanwhile, LPIPS is designed to match human perception and yield better scores for images with higher level of coherence.
Note that the recorded time cost for PBO does not include view selection or is not for texturing the entire proxy although it is relatively long.

\begin{figure}[!tbh]
	\centering
	\includegraphics[width=\linewidth]{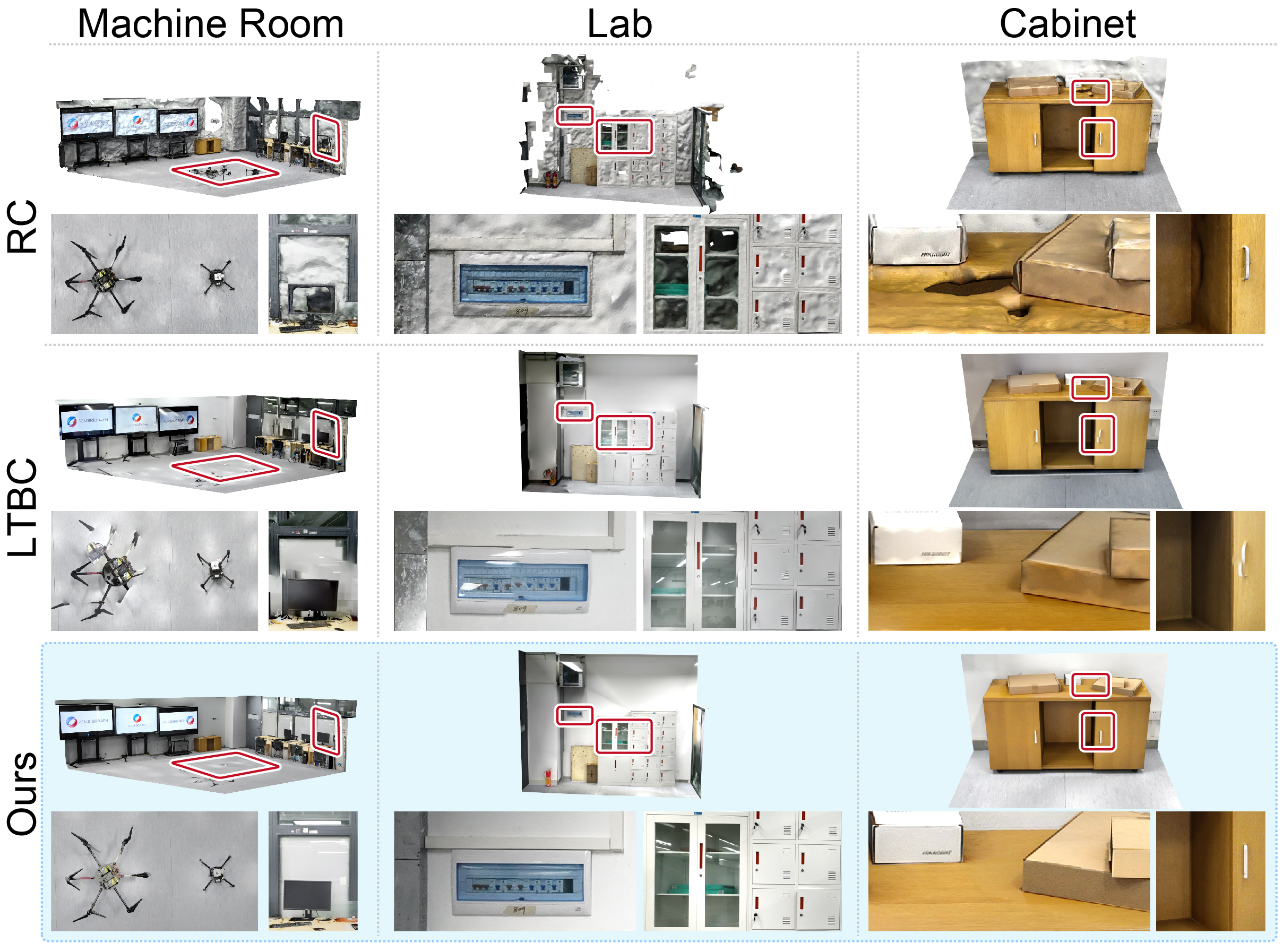}
	\caption{\wdx{Visual comparison against RC, LTBC and ours on two indoor scenes and a cabinet.}}
	\label{fig:more-cases}
\end{figure}

\subsection{Analysis of Generalization Ability}

In this subsection, we discuss the results of our method applied to more diverse scenes. First, we conduct experiments on several challenging urban scenes that contain curve surfaces, facades with non-linear features, and facades with a lot of reflection, as shown in Fig.~\ref{fig:challenging-cases}. Then we present the texturing results on two indoor scenes and one man-made object in Fig.~\ref{fig:more-cases}, which are different scenarios from buildings. Note there exists only one fire extinguisher in the Lab scene. Although it has been moved during the photos collection process, our method still perfectly keeps its location where it has been placed in most photos.

\begin{figure}
	\centering
	\includegraphics[width=\linewidth]{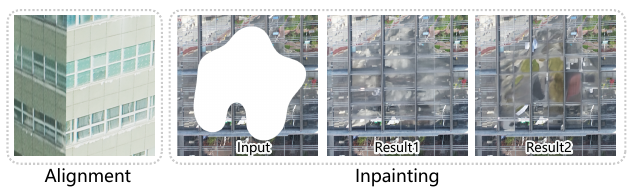}
	\caption{Failure examples: inter-planar misalignment among two adjacent facades (left), and inpainting results of a reflective facade \jw{with large glass panels} (right).}
	\label{fig:limitations}
\end{figure}

To better understand the superiority of our TwinTex, we conduct comparisons with previous methods as well. From the figures we can see that our approach can still generate high-fidelity textures for the challenging buildings, and can be naturally extended to texture coarse piece-wise planar models in other scenarios.
The visual comparisons further suggest that the textures generated by our approach have several significant advantages over previous methods: i) closer to real photos with perspective consistency and harmonic illumination, ii) contain rarely border misalignment, distortion, blurring and seaming artifacts resulting from inaccurate camera parameters, iii) contain rarely ghosting effects resulting from dynamic instances, iv) preserve geometric details better, and v) fill the missing regions with geometrically and semantically consistent contents.

\section{Conclusion and Future Work}

In this work, we proposed a plane-based and geometry-aware texture generation method. Our target is to produce high-resolution texture maps for piece-wise planar architectural proxy models which were abstracted from dense 3D reconstructions. Our main technical contributions consist in:
i) A greedy algorithm to select a set of views that are most suitable for texturing each planar shape; Both perspective and photometric qualities are considered in this algorithm.
ii) An image stitching methodology that can stitch the selected views to an enlarged texture map, preserving the local and global linear geometric primitives based on an adaptive-mesh data structure;
iii) An improved diffusion probabilistic model that fine-tuned with our created TwinTexSet to inpaint unobserved texture region with semantically and geometrically coherent pixels;
iv) A customized texturing tool plugged in a widely used graphical engine. Experimental results show that our algorithm outperforms state-of-the-art facet-based texturing methods in the generation of realistic and high-resolution texture maps in a reasonable time.

\textbf{Limitations and future work.}
Our system still has several limitations.
First, although we significantly refined the extracted line segments, the line-guided scheme still relies on the quality of the line extraction algorithm. One possible solution is to consider camera information as well or make use of reconstructed 3D lines.
Second, our plane-based texture generation method processes each planar shape independently, which does not take their mutual relationship into consideration\ignore{ in a more global manner}. Fig.~\ref{fig:limitations} (left) shows such an example where global alignment of linear features between adjacent planes is not achieved. Note we here choose a zoom-in view with the most obvious misalignment among given buildings to clearly show this issue. The other examples may also exhibit tiny misalignments.
Third, we may fail to successfully perform inpainting for a reflected facade which lacks of coherent structures and clear content, making it even hard for a human to infer the missing region, see Fig.~\ref{fig:limitations} (right).
Finally, the processing time can be further improved with parallel processing. We will incorporate these issues in our future pipeline.

\section*{Acknowledgments}
We thank the reviewers for their constructive comments. This work was supported in parts by NSFC (U21B2023, U2001206, U22B2034, 62172416, 62302313), DEGP Innovation Team (2022KCXTD025), and Shenzhen Science and Technology Program (KQTD202108110900440 03, RCJC20200714114435012, JCYJ20210324120213036).

\bibliographystyle{ACM-Reference-Format}
\bibliography{TexRef}
\end{document}